\documentclass[aps,prd,twocolumn,showkeys,floats,showpacs, superscriptaddress,nofootinbib]{revtex4-1}

\usepackage{graphicx}

\usepackage{bm}
\usepackage{amsmath,amssymb}
\usepackage{aas_macros}
\usepackage{natbib}
\usepackage[colorlinks=true]{hyperref}

\begin{document}
\title{Violation of the equivalence principle from light scalar dark matter}

\author{Aur\'elien Hees}
\email{aurelien.hees@obpsm.fr}
\affiliation{SYRTE, Observatoire de Paris, Universit\'e PSL, CNRS, Sorbonne Universit\'e, LNE, 61 avenue de l'Observatoire 75014 Paris, France}

\author{Olivier Minazzoli}
\affiliation{Centre Scientifique de Monaco, 8 Quai Antoine 1er, 98000 Monaco, Monaco}
\affiliation{Laboratoire Artemis, Universit\'e C\^ote d'Azur, CNRS, Observatoire C\^ote d'Azur, BP4229, 06304, Nice Cedex 4, France}

\author{Etienne Savalle}
\affiliation{SYRTE, Observatoire de Paris, Universit\'e PSL, CNRS, Sorbonne Universit\'e, LNE, 61 avenue de l'Observatoire 75014 Paris, France}

\author{Yevgeny V. Stadnik}
\affiliation{Helmholtz Institute Mainz, Johannes Gutenberg University, 55099 Mainz, Germany}

\author{Peter Wolf}
\affiliation{SYRTE, Observatoire de Paris, Universit\'e PSL, CNRS, Sorbonne Universit\'e, LNE, 61 avenue de l'Observatoire 75014 Paris, France}

\begin{abstract}
	In this paper, we study the local observational consequences of a violation of the Einstein Equivalence Principle induced by models of light scalar Dark Matter (DM). We focus on two different models where the scalar field couples linearly or quadratically to the standard model of matter fields. For both these cases, we derive the solutions of the scalar field. 
	
	We also derive from first principles the expressions for two types of observables: (i) the local comparison of two atomic sensors that are differently sensitive to the constants of Nature and (ii) the local differential acceleration between two test masses with different compositions. For the linear coupling, we recover that the signatures induced by DM on both observables are the sum of harmonic and Yukawa terms. For the quadratic coupling on the other hand, the signatures derived for both types of observables turn out to be the sum of a time-independent term and a harmonic oscillation, whose amplitudes both depend on the position. Such behavior is new and can make experiments in space more sensitive than terrestrial ones. Besides this, the observables present some interesting nonlinear behaviors that are due to the  amplification or to the screening of the scalar field, depending on the parameters of the theory, and on the compactness of the source of the gravitational field.
	
	Finally,  we infer the various limits on the DM coupling parameters by using existing frequency comparisons on the one hand and tests of the universality of free fall on the ground (torsion balances) or in space (MICROSCOPE mission) on the other hand. We show that in the quadratic case, so-called natural parameters are still allowed by observations.
\end{abstract}


\maketitle

\clearpage

\section{Introduction}
While thoroughly tested experimentally \cite{will:2014la}, the theory of General Relativity (GR) is currently challenged by galactic and cosmological observations that may require the introduction of the so-called Dark Matter (DM) and Dark Energy (see e.g. \cite{debono:2016zr,famaey:2012fk}). Besides this, several theoretical developments of a quantum theory of gravitation and of a theory that would unify GR with the standard model of particle physics also challenge GR. Motivated by the unsuccessful searches for a Dark Matter particle at high energy, models of light scalar DM have recently gained a lot of attention in the scientific community (see e.g. \cite{weinberg:1978rc,preskill:1983fv,hu:2000aa,piazza:2010aa,khmelnitsky:2014aa,porayko:2014aa,schive:2014aa,beyer:2014aa,arvanitaki:2015qy,stadnik:2015aa,stadnik:2015yu,graham:2016ab,arvanitaki:2016qv,stadnik:2016aa,stadnik:2016kq,marsh:2016aa,urena-lopez:2016aa,calabrese:2016aa,blas:2017aa,bernal:2017aa,hui:2017aa,abel:2017aa,bernal:2018aa} and references therein). In those models, a light scalar field is introduced in addition to the standard space-time metric and to the standard model fields. Such scalar fields are also ubiquitous in theories with more than 4 dimensions, and in particular in string theory with the dilaton and the moduli fields \cite{green:1988oa,damour:1994fk,*damour:1994uq,gasperini:2002kx,damour:2002ys}. 

In the simplest models, the scalar field has a regular quadratic kinetic term and a standard quadratic potential from which it gets its mass. In the most general scenario, this scalar field couples nonuniversally to the standard model fields, which leads to a violation of the Einstein Equivalence Principle (EEP). Such models have been shown to produce nice galactic and cosmological predictions for very low masses of the scalar field ranging from $10^{-24}$ to $10^{-22}$ eV~\cite{hu:2000aa,beyer:2014aa,schive:2014aa,urena-lopez:2016aa,calabrese:2016aa,marsh:2016aa,bernal:2017aa,hui:2017aa,bernal:2018aa}. Because of the high occupation numbers in galactic halos, the scalar field can be treated as a classical field for masses $\ll$ eV~\cite{hu:2000aa,stadnik:2016kq}.
 
A convenient microscopic modeling for the coupling between the scalar field and standard matter has been introduced by Damour and Donoghue \cite{damour:2010ve,damour:2010zr}. In this seminal work, the scalar matter coupling is assumed to be linear in terms of the scalar field, whereas a quadratic generalization is also often used \cite{stadnik:2015aa,stadnik:2015yu,stadnik:2016aa,stadnik:2018aa,kalaydzhyan:2017aa}. The main property of this model lies in the fact that the constants of Nature (like the electromagnetic fine structure constant $\alpha_\textrm{EM}$, the masses of the fermions or the quantum chromodynamics (QCD) energy scale $\Lambda_3$) become directly linearly or quadratically dependent on the scalar field \cite{damour:2010ve,damour:2010zr}. 

If the scalar field is massive and if its mass is larger than $m_\varphi \gg \hbar H/c^2\sim 1.5\times 10^{-33}$ eV/c$^2$ (where $H$ is the Hubble constant)~\cite{arvanitaki:2015qy}, it will oscillate in time at its Compton frequency, causing the constants of Nature to oscillate as well. This is one characteristic of a violation of the EEP~\cite{will:2014la,uzan:2011vn} that can be searched for with various local experiments, in particular, with atomic sensors \cite{safronova:2018aa,stadnik:2018aa}.  Another consequence induced by a violation of the EEP is a violation of the Universality of Free Fall (UFF), which can also be constrained by different measurements.

In this paper, we first recover that, in the case of a linear coupling between the scalar field and standard matter, the scalar field is made of two contributions: an oscillating solution which can be identified as DM, and a Yukawa solution generated by standard matter. We recover that this leads to two different types of signatures that can be searched for in measurements: (i) an oscillatory signature, for which atomic sensors are particularly adapted, and (ii) a fifth force, for which UFF measurements are particularly powerful. However, we show that this situation is dramatically different in the case of a quadratic coupling for which no classical Yukawa solution for the scalar field is allowed \cite{de-pirey-saint-Alby:2017aa}. In that case, the scalar field exhibits a harmonic behavior whose amplitude can be enhanced or screened by standard matter, a mechanism that is similar to the scalarization \cite{damour:1993vn,*damour:1996uq}. This new behavior is fundamentally different from the one arising with a linear coupling.

We then use several existing measurements to constrain the various coefficients that parametrize the coupling between the scalar field and standard matter for both the linear and quadratic couplings. The measurements used in this paper are the ones from torsion balances \cite{smith:1999aa,schlamminger:2008zr,wagner:2012fk} and from the MICROSCOPE space experiment \cite{touboul:2017aa}, as well as from local comparisons of atomic clocks \cite{van-tilburg:2015fj,hees:2016uq}.
The constraints obtained in the linear case summarize existing results while most of the constraints obtained in the quadratic case are new.

First, in Sec.~\ref{sec:action_field} we thoroughly present the scalar DM model considered in this paper, as well as the microscopic interactions between the scalar field and matter. In Sec.~\ref{sec:matter_modeling}, we detail what are the macroscopic modelings of observables that derive from the microscopic Lagrangian introduced in Sec.~\ref{sec:action_field}. We then discuss the solution for the scalar field around a spherical body in both the linear and quadratic cases in Sec.~\ref{sec:phi} while the detailed calculations are developed in Appendix~\ref{app:sol}. The solutions for the scalar field are then used to derive the observable signatures induced by a violation of the equivalence principle in Sec.~\ref{sec:obs}. Finally, the constraints on the various parameters that are obtained by using different experimental data are presented and discussed in Sec.~\ref{sec:constraints}. 

\section{Action and field equations}\label{sec:action_field}

In the present paper, we consider the following action:
\begin{align}
	S&=\frac{1}{c}\int d^4x \frac{\sqrt{-g}}{2\kappa}\left[R-2g^{\mu\nu}\partial_\mu\varphi\partial_\nu\varphi-V(\varphi)\right] \nonumber\\
	&+ \frac{1}{c}\int d^4x \sqrt{-g}\Bigg[ \mathcal L_\textrm{SM}[g_{\mu\nu},\Psi_i]  + \mathcal L_\textrm{int}[g_{\mu\nu},\varphi,\Psi_i]  \Bigg]\, , \label{eq:action}
\end{align}
where $\kappa=8\pi G/c^4$, $R$ is the Ricci scalar of the space-time metric $g_{\mu\nu}$, $\varphi$ is a dimensionless scalar field (note that a dimensional scalar field $\phi$ is sometimes used, see Appendix~\ref{app:conventions}), $\mathcal L_\textrm{SM}$ is the Lagrangian density of the Standard Model of particles depending on the standard model fields $\Psi_i$, and $\mathcal L_\textrm{int}$ parametrizes the interaction between matter and the scalar field.  In this communication, we consider linear and quadratic couplings between matter and the scalar field. Following  \cite{damour:2010zr,damour:2010ve,stadnik:2015yu}, we consider two phenomenological microscopic modelings for the coupling between the scalar and matter fields: (i) a linear coupling parametrized by
\begin{subequations}\label{eq:Lint}
\begin{align}\label{eq:Lint_lin}
 \mathcal L_\textrm{int}^{(1)}&=	\varphi  \Bigg[\frac{d^{(1)}_e}{4\mu_0}F^2-\frac{d_g^{(1)}\beta_3}{2g_3}\left(F^A\right)^2\\
& \quad -\sum_{i=e,u,d}\Big(d_{m_i}^{(1)}+\gamma_{m_j}d_g^{(1)}\Big)m_i\bar\psi_i\psi_i\Bigg] \, , \nonumber
\end{align}
and (ii) a quadratic coupling parametrized by
\begin{align}\label{eq:Lint_quad}
 \mathcal L_\textrm{int}^{(2)}&=	\frac{\varphi^2}{2}  \Bigg[\frac{d^{(2)}_e}{4\mu_0}F^2-\frac{d_g^{(2)}\beta_3}{2g_3}\left(F^A\right)^2\\
& \quad -\sum_{i=e,u,d}\Big(d_{m_i}^{(2)}+\gamma_{m_j}d_g^{(2)}\Big)m_i\bar\psi_i\psi_i\Bigg] \, , \nonumber
\end{align}
\end{subequations}
with $F_{\mu\nu}$ being the standard electromagnetic Faraday tensor, $\mu_0$ the magnetic permeability, $F^A_{\mu\nu}$ the gluon strength tensor, $g_3$ the QCD gauge coupling, $\beta_3$ the $\beta$ function for the running of $g_3$, $m_j$ the mass of the fermions (electron and light quarks \footnote{Following the most recent literature \cite{dzuba:2008uq}, we do not take into account the effects of the strange quark, although they have been estimated in the past for atomic clocks measurements \cite{flambaum:2004pr, flambaum:2006pr}.}), $\gamma_{m_j}$ the anomalous dimension giving the energy running of the masses of the QCD coupled fermions and $\psi_j$ the fermion spinors.  The constants $d_j^{(i)}$ characterize the interaction between the scalar field $\varphi$ and the different matter sectors. Note that another convention for the coupling coefficients is sometimes considered using dimensional $\Lambda_i$ coupling constants (see Appendix~\ref{app:conventions}) and that some authors~\cite{olive:2008fk} consider a more general case of coupling which takes the form of $d_j(\varphi-\varphi_j)^2$ and corresponds to a linear combination of both linear and quadratic Lagrangians. This Lagrangian leads to the following effective dependency of five constants of Nature
\begin{subequations}\label{eq:constants}
	\begin{align}
		\alpha_\textrm{EM}(\varphi)&=\alpha_\textrm{EM}\left(1+d^{(i)}_e\frac{\varphi^i}{i}\right) \, , \\
		m_j(\varphi)&=m_j\left(1+d^{(i)}_{m_j}\frac{\varphi^i}{i}\right) \quad \textrm{for } j=e,u,d\,  \\
		\Lambda_3(\varphi)&= \Lambda_3\left(1+d^{(i)}_{g}\frac{\varphi^i}{i}\right) \, ,
	\end{align}
\end{subequations}
where $\alpha_\textrm{EM}$ is the electromagnetic fine structure constant, $m_j$ are the masses of the fermions  (the electron and the up, down and strange quarks), $\Lambda_3$ is the QCD mass scale $\Lambda_3$ and the superscripts $^{(i)}$ indicate the type of coupling considered (linear for $i=1$ and quadratic for $i=2$). Note that, following Damour and Donoghue \cite{damour:2010zr,damour:2010ve}, we introduce the mean quark mass $\hat m=\left(m_u+m_d\right)/2$ and the difference of the quark masses $\delta m=m_d-m_u$ \footnote{Besides, note that the assumption $m_u=m_d$ is often used in the nuclear physics calculations that must be used in the interpretation of the atomic sensors phenomenology \cite{flambaum:2004pr, flambaum:2006pr,dzuba:2008uq}. Therefore, in the present paper, one implicitly has $\hat m = m_u = m_d$ when considering clock measurements.}, which depend also on the scalar field through
\begin{subequations}
\begin{align}
	\hat m(\varphi)&=\hat m\left(1+d^{(i)}_{\hat m}\frac{\varphi^i}{i}\right)\, \\
	\delta m(\varphi)&=\delta m\left(1+d^{(i)}_{\delta m}\frac{\varphi^i}{i}\right)\, ,
\end{align}
\end{subequations}
with
$$
 d^{(i)}_{\hat m}=\frac{m_u d^{(i)}_{m_u}+m_d d^{(i)}_{m_d}}{m_u+m_d}\, , \, d^{(i)}_{\delta m}=\frac{m_d d^{(i)}_{m_d}-m_u d^{(i)}_{m_u}}{m_d-m_u}\, .
$$
We also consider a quadratic scalar potential 
\begin{equation}
	V(\varphi)=2\frac{c^2}{\hbar^2}m_\varphi^2\varphi^2\, ,
\end{equation}
where $m_\varphi$ has the dimension of a mass.

The field equations deriving from action (\ref{eq:action}) are
\begin{subequations}
	\begin{align}
			R_{\mu\nu}&=\kappa \left[T_{\mu\nu}-\frac{1}{2}g_{\mu\nu}T\right] +2\partial_\mu\varphi\partial_\nu\varphi +\frac{1}{2}g_{\mu\nu}V(\varphi)\, , \label{eq:einstein}\\
	\Box \varphi&=-\frac{\kappa}{2}\sigma  +\frac{V'(\varphi)}{4}  \, ,\label{eq:scalar_field}
	\end{align}
	with
	\begin{align}
		T_{\mu\nu}&=-\frac{2}{\sqrt{-g}}\frac{\delta \sqrt{-g}\mathcal L_\textrm{mat}}{\delta g^{\mu\nu}} \, , \\
		\sigma&=\frac{1}{\sqrt{-g}}\frac{\delta \sqrt{-g}\mathcal L_\textrm{mat}}{\delta\varphi} =\frac{\partial \mathcal L_\textrm{int}}{\partial \varphi}\, .
	\end{align}
\end{subequations}

\section{Matter and clock modeling}\label{sec:matter_modeling}
\subsection{Test masses}
Damour and Donoghue have shown that the action used to model matter at the microscopic level including the scalar field interaction from Eq.~(\ref{eq:Lint}) can phenomenologically be replaced at the macroscopic level by a standard point mass action
\begin{equation}\label{eq:smat}
		S_\textrm{mat}[g_{\mu\nu},\varphi,\Psi_i]=-c^2\sum_A \int_A d\tau~ m_A(\varphi) \, ,
\end{equation}
where $d\tau$ is the proper time interval defined by $c^2d\tau^2=-g_{\alpha\beta}dx^\alpha dx^\beta$. Each mass $A$ has its own composition such that the function $m_A(\varphi)$ will be different. The effects produced by the coupling of the dilaton to matter are encoded in the coupling function
\begin{equation}\label{eq:alpha}
	\alpha_A(\varphi)=\frac{\partial \ln m_A(\varphi)}{\partial \varphi}\, .
\end{equation}
Damour and Donoghue \cite{damour:2010zr,damour:2010ve} have derived a semianalytical expression for the coupling $\alpha_A(\varphi)$. These expressions are given in Appendix~\ref{app:dilatonic_charges}. It is convenient to separate these couplings into composition-dependent and -independent parts, which leads to (see Appendix~\ref{app:dilatonic_charges} for further details)
\begin{subequations}
	\begin{eqnarray}
		\alpha_A^{(1)} &= d_g^{*(1)}+\bar \alpha_A^{(1)}& \quad \textrm{for a linear coupling}\, \\
						&= \tilde \alpha^{(1)}_A \, ,\nonumber \\
		\alpha_A^{(2)} &= d_g^{*(2)}\varphi+\bar \alpha_A^{(2)} \varphi&\quad \textrm{for a quad. coupling} \\
						&= \tilde \alpha^{(2)}_A \varphi \, , \nonumber
	\end{eqnarray}	
\end{subequations}
where we introduce
\begin{equation}\label{eq:tilde_alpha}
	\tilde \alpha_A^{(i)}=d_g^{*(i)}+\bar \alpha_A^{(i)}\, .
\end{equation}
The universal part of the coupling between matter and the scalar field $d_g^{*(i)}$ is expressed in terms of the fundamental scalar field and matter coupling constants that enter the interaction part of the Lagrangian from Eq.~(\ref{eq:Lint}) as (see \cite{damour:2010zr,damour:2010ve} and Appendix~\ref{app:dilatonic_charges})
\begin{align}
	d_g^{*(i)}=&d_g^{(i)}+0.093\Big(d_{\hat m}^{(i)}-d_g^{(i)}\Big)+2.75\times 10^{-4}\Big(d_{ m_e}^{(i)}-d_g^{(i)}\Big)\nonumber\\
	&\qquad\quad+2.7\times 10^{-4}\, d_e^{(i)}\, ,
\end{align}
On the other hand, the composition-dependent part of $\alpha_A$ is given by
\begin{align}
	\bar\alpha^{(i)}=&\left[Q'_{\hat m}\right]_A\Big(d_{\hat m}^{(i)}-d_g^{(i)}\Big) +\left[Q'_{ m_e}\right]_A\Big(d_{ m_e}^{(i)}-d_g^{(i)}\Big) \nonumber\\
	&+ \left[Q'_{e}\right]_Ad_e^{(i)}+\left[Q'_{\delta m}\right]_A\Big(d_{\delta m}^{(i)}-d_g^{(i)}\Big)\, ,\label{eq:bar_alpha}
\end{align}
where the coefficients $\left[Q'_j\right]_A$ are the dilatonic charges for the body $A$. The values of these coefficients depend only on the composition of each body. Semiempirical expressions for these coefficients have been derived in \cite{damour:2010zr,damour:2010ve}, their expressions are given in Appendix~\ref{app:dilatonic_charges}, Eq.~(\ref{eq:Q'}) while the important values for the present work are summarized in Table~\ref{tab:dilaton_coef}. 

\begin{table}[htb]
\caption{Values of the dilatonic charges for different materials.}
\label{tab:dilaton_coef} 
\centering
\begin{tabular}{c c c c c c}
\hline
Material & $-Q'_{\hat m}$     & $Q'_e$             & $-Q'_{m_e}$        & $Q'_{\delta m}$\\
         & $[\times 10^{-3}]$ & $[\times 10^{-3}]$ & $[\times 10^{-5}]$ & $[\times 10^{-4}]$ \\
\hline
 H/He [70:30]       & $45.51$            & $0.36$  &  $-18.9$ &  $-11.7$\\
Fe                  & $\phantom{0}9.94$  & $2.32$  &  $1.89$  &  $1.17$\\
SiO$_2$             & $13.70$            & $1.26$  &  $0.027$ &   $0.02$    \\ \hline
Be                  & $17.64$            & $0.45$  &  $3.05$ &  $1.91$\\
Al                  & $12.30$            & $1.47$  &  $1.00$ &  $0.62$\\
Ti                  & $10.42$            & $2.01$  &  $2.24$ &  $1.38$\\ 
$^{238}$U           & $\phantom{0}7.63$  & $4.28$  &  $6.24$ &  $3.86$\\ 
Cu                  & $\phantom{0}9.63$  & $2.46$  &  $2.18$ &  $1.35$\\ 
Pb                  & $\phantom{0}7.73$  & $4.06$  &  $5.82$ &  $3.60$\\
\hline
Pt/Rh [90:10]       & $\phantom{0}7.83$  & $3.92$  &  $5.30$ &  $3.28$\\
Ti/Al/V [90:6:4]    & $10.52$            & $1.98$  &  $2.17$ &  $1.34$\\
\hline
\end{tabular}
\end{table}

\subsection{Atomic clocks}
Atomic clocks are sensitive to a hypothetical variation of the constants of Nature from Eq.~(\ref{eq:constants}). A standard way to parametrize a possible variation of any atomic frequency $X$ to variations of the constants of Nature is to use the following parametrization (see e.g. \cite{flambaum:2004fk,guena:2012ys})
\begin{equation}\label{eq:clock1}
	d \ln X=\left[k_\alpha\right]_X d\ln \alpha_\textrm{EM} +\left[ k_\mu\right]_X d\ln \mu + \left[k_q\right]_X d\ln m_q/\Lambda_3 \, ,
\end{equation}
where $\mu=m_e/m_p$ is the ratio of the electron mass over the proton mass, $m_q$ is the mass of the light quarks (assumed to be equal),  and the $k_i$'s are the sensitivity coefficients of the specific transition $X$. The atomic and nuclear calculations to derive these sensitivity coefficients have been achieved in \cite{prestage:1995pl,flambaum:2004fk,flambaum:2006pr,dzuba:2008uq}, and the obtained numerical values can be found in Table I of Ref. \cite{guena:2012ys}.

While the parametrization in Eq. (\ref{eq:clock1}) is widely used, another equivalent parametrization is useful since it is closer to the form of the interaction Lagrangian from Eq.~(\ref{eq:Lint})
\begin{eqnarray}
		d \ln X&=&\left[k_\alpha\right]_X d\ln \alpha_\textrm{EM} + \left[k_\mu\right]_X d\ln m_e/\Lambda_3\nonumber \\
&&+ \left[k'_q\right]_X d\ln m_q/\Lambda_3 \, , \label{eq:clock}
\end{eqnarray}
with $k'_q=k_q-0.049$ \cite{yang:2016al}. These sensitivity coefficients play a role equivalent to those of the dilatonic charges introduced in the previous section. 

The coupling of the scalar field to a clock working on the transition $X$ is then encoded in the coupling function $\kappa_X$  which is defined by
\begin{equation}\label{eq:clock_def}
	d \ln X=\kappa^{(i)}_X d\left( \varphi^i\right) \, ,
\end{equation}
and can be expressed as
\begin{align}
	\kappa^{(i)}_X =& \frac{1}{i}\left[k_\alpha\right]_X d^{(i)}_e + \frac{1}{i}\left[k_\mu\right]_X\left(d^{(i)}_{m_e}-d^{(i)}_{g}\right) \nonumber\\
	&\qquad +\frac{1}{i}\left[k'_q\right]_X \left(d^{(i)}_{\hat m}-d^{(i)}_{g}\right) \, . \label{eq:kappa}
\end{align}

\section{Solutions for the scalar field}\label{sec:phi}
The space-time evolution of the scalar field depends on the distribution of matter. In this manuscript, we will consider spherically symmetric extended bodies that will be characterized by a radius $R_A$ and by a constant matter density $\rho_A$. The case of a two-layer spherical body is also considered in Appendix \ref{app:sol}.

At first order, we model standard matter as a pressureless perfect fluid whose stress-energy tensor is given by $T^{\mu\nu}=c^2 \rho u^\mu u^\nu$, where $\rho$ is the matter density and $u^\nu$ the 4-velocity of the fluid \footnote{Corrections due to the pressure will arise at the post-Newtonian order and can safely be neglected here.}. For this matter modeling, the source term in the Klein-Gordon equation~(\ref{eq:scalar_field}) is written as
\begin{equation}
	\sigma = -\alpha(\varphi)\rho c^2 \, ,
\end{equation}
where $\alpha$ is given by Eq.~(\ref{eq:alpha}).

At the Minkowskian order, the equation for the scalar field (\ref{eq:scalar_field}) is
\begin{equation}
	\frac{1}{c^2}\ddot \varphi(t,\bm x) - \Delta \varphi(t,\bm x) =  -\frac{4\pi G }{c^2}\alpha_A(\varphi)\rho_A(\bm x)-\frac{c^2m_\varphi^2}{\hbar^2}\varphi(t,\bm x)\, ,\label{eq:KG}
\end{equation}
where the dot denotes a derivative with respect to the coordinate time $t$ and $\Delta$ is the 3-dimensional flat Laplacian. In this equation, we have neglected terms that are of the order of $\mathcal O(\left|h_{\mu\nu}\right|)$ (with $h_{\mu\nu}=g_{\mu\nu}-\eta_{\mu\nu}$). Indeed, a linearized version of the Einstein equation (\ref{eq:einstein}) shows that the metric will be generated by sources that will contribute as $\sim \frac{GM_A}{c^2r}\ll1$ and by terms that are proportional to $\varphi_0^2$ ($\varphi_0$ being the typical amplitude of the scalar field). If the scalar field is associated with the local galactic DM abundance, one can show that $\varphi_0\sim 7\times 10^{-31} \textrm{ eV}/m_\varphi$ \cite{van-tilburg:2015fj,hees:2016uq} which shows that $\varphi_0^2\ll1$ for scalar field masses above $10^{-30}$ eV. Under this assumption, the space-time behavior of the scalar field will be governed by Eq.~(\ref{eq:KG}) whose solution will be given in this section.
Nevertheless, the explicit limit at which this assumption breaks down has been carefully taken into account when deriving the constraints on the parameters $d_i$ in Sec.~\ref{sec:constraints}.

\subsection{Linear coupling}
In the case of a linear coupling, the function $\alpha_A(\varphi)=\tilde\alpha_A^{(1)}$ appearing in Eq.~(\ref{eq:KG}) is independent of the scalar field and the general solution is a sum of free waves and a Yukawa-type scalar field generated by the central body. Details about the derivation of the results are given in Appendix \ref{app:sol}. The general expression of the scalar field is given by
\begin{align}\label{eq:phi_1}
	\varphi^{(1)}(t,\bm x) &=  \varphi_0 \cos \left(\bm k .\bm x -\omega t+\delta\right) - s^{(1)}_A \frac{GM_A}{c^2r}e^{-r/\lambda_\varphi}\, , 
\end{align}
where  $\left|\bm k\right|^2 + c^2 m_\varphi^2/\hbar^2=\omega^2/c^2$ and
\begin{equation}
	\lambda_\varphi= \frac{\hbar}{c m_\varphi} \, ,
\end{equation}
is the reduced Compton wavelength of the scalar field. 
The constant $s^{(1)}_A$ is the effective scalar charge of the extended body and is given by
\begin{align}\label{eq:s1}
	s^{(1)}_A&= \tilde \alpha^{(1)}_A I \left (\frac{R_A}{\lambda_\varphi} \right)\, ,
\end{align} 
with the function $I(x)$ given by
$$
	I(x)=3 \frac{x\cosh x-\sinh x }{x^3}\, .
$$
Note that this result, valid only for a homogeneous sphere, is generalized to a two-layer sphere in Appendix \ref{app:sol}. The only difference is related to the expression of the effective scalar charge $s_A$ which would be given by Eq.~(\ref{eq:s1_app}).

\subsection{Quadratic coupling}
\label{sec:sol_quad}
\begin{figure}[hbt]
\begin{center}
\includegraphics[width=0.4\textwidth]{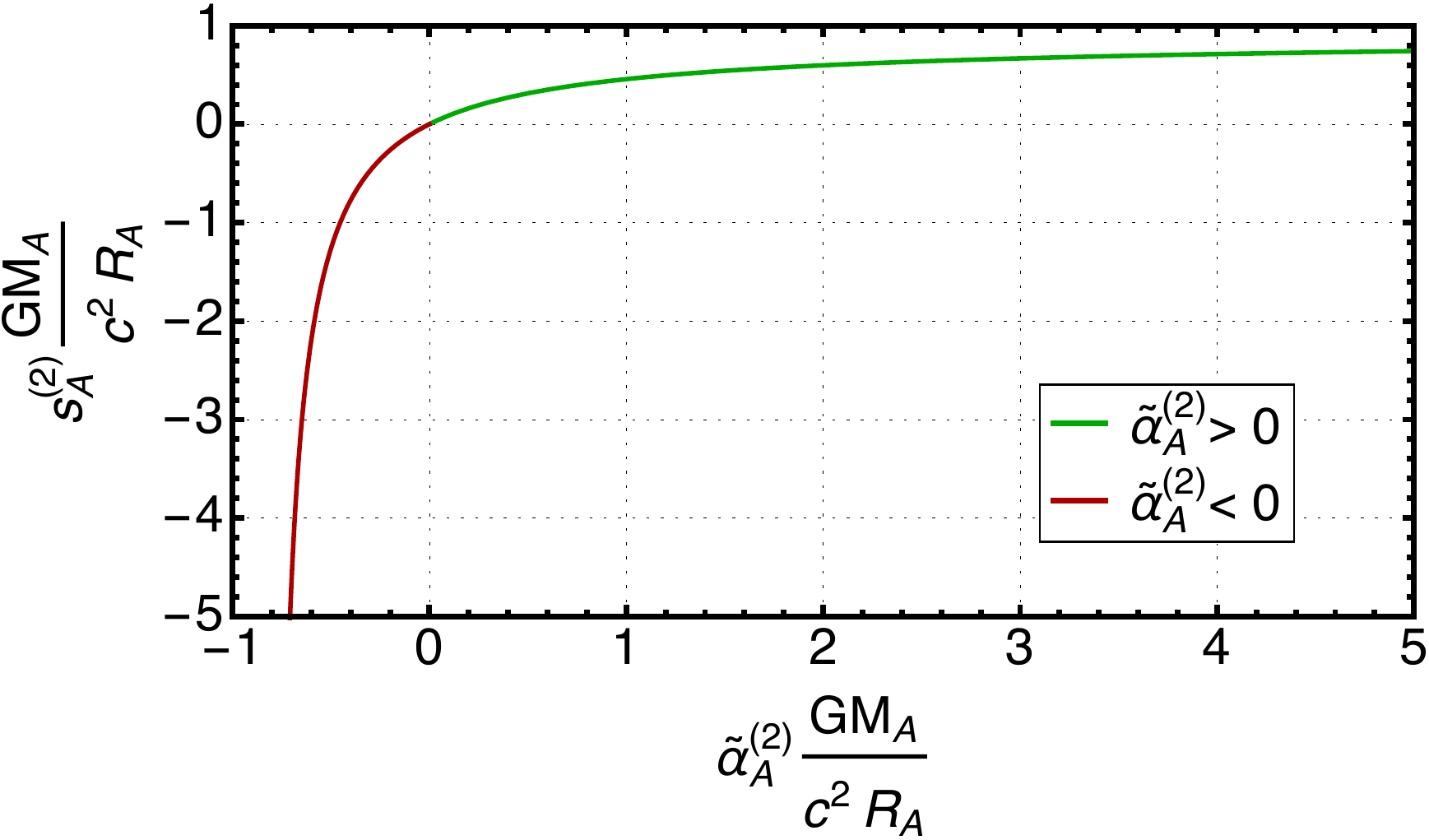}
\end{center}
\caption{Evolution of the effective scalar charge $s^{(2)}_A$ that appears in the solution of the scalar field from Eq. (\ref{eq:phi_2}) as a function of $\varepsilon_A$ from Eq.~(\ref{eq:eps}). For large positive values of $\varepsilon_A$, a deamplification mechanism occurs and the scalar field at the surface of the body tends to vanish. On the other hand, for negative values of $\varepsilon_A$, the scalar field is amplified, which leads to nonperturbative effects.}
\label{fig:scalar_charge}
\end{figure}

In the case of a quadratic coupling, the function $\alpha_A(\varphi)=\tilde\alpha_A^{(2)} \varphi$ that appears in the Klein-Gordon equation (\ref{eq:KG}) is now linear in $\varphi$. This linear dependency changes drastically the form of the solution. In particular, in the classical limit, it is easy to show that there exists no static solution beyond the trivial one\footnote{Quantum one-loop corrections are expected to produce an additional $1/r^3$ potential at distances $2m_\varphi r\ll 1$, see e.g. \cite{olive:2008fk}.}. The time-dependent solution contains several modes, but only one is nonvanishing at infinity and can be interpreted as DM (see Appendix~\ref{app:sol} for details). Its expression is given by
\begin{equation}\label{eq:phi_2}
	\varphi^{(2)}(t,\bm x)=\varphi_0 \cos \left(\frac{m_\varphi c^2}{\hbar}t+\delta\right)\left[1-s^{(2)}_A\frac{GM_A}{c^2r}\right]\, ,
\end{equation}
with the effective scalar charge
\begin{equation} \label{eq:sa}
	s^{(2)}_A=\tilde \alpha_A^{(2)} J_{\textrm{sign}[\tilde \alpha_A^{(2)}]}\left(\sqrt{3 \left|\tilde \alpha_A^{(2)} \right| \frac{GM_A}{c^2R_A}}\right) \, ,
\end{equation}
which depends on the sign of $\tilde \alpha_A^{(2)}$ through
\begin{subequations} \label{eq:Js}
	\begin{align}
		J_+(x)&=3 \frac{x-\tanh x}{x^3} \, , \\
		J_-(x)&=3 \frac{\tan x-x}{x^3} \, .
	\end{align}
\end{subequations}
$J_+$ corresponds to the cases such that $\tilde\alpha_A^{(2)} >0$ while $J_-$ corresponds to the cases such that $\tilde\alpha_A^{(2)} <0$. In the limit of weak gravitational fields and small coupling constants (i.e. $x\ll1$), $J_\pm(x)\approx1$ and $s^{(2)}_A\approx \tilde \alpha_A^{(2)}$. In this case, note that the expression of the scalar field is similar to the one derived in \cite{de-pirey-saint-Alby:2017aa}. The behavior of the scalar field around a body $A$ -- through the effective scalar charge $s^{(2)}_A$ -- depends only on the dimensionless parameter 
\begin{equation}\label{eq:eps}
	\varepsilon_A = \tilde \alpha_A^{(2)} \frac{GM_A}{c^2R_A}\, ,
\end{equation}
as illustrated in Fig.~\ref{fig:scalar_charge}.

\begin{figure}[hbt]
\begin{center}
\includegraphics[width=0.4\textwidth]{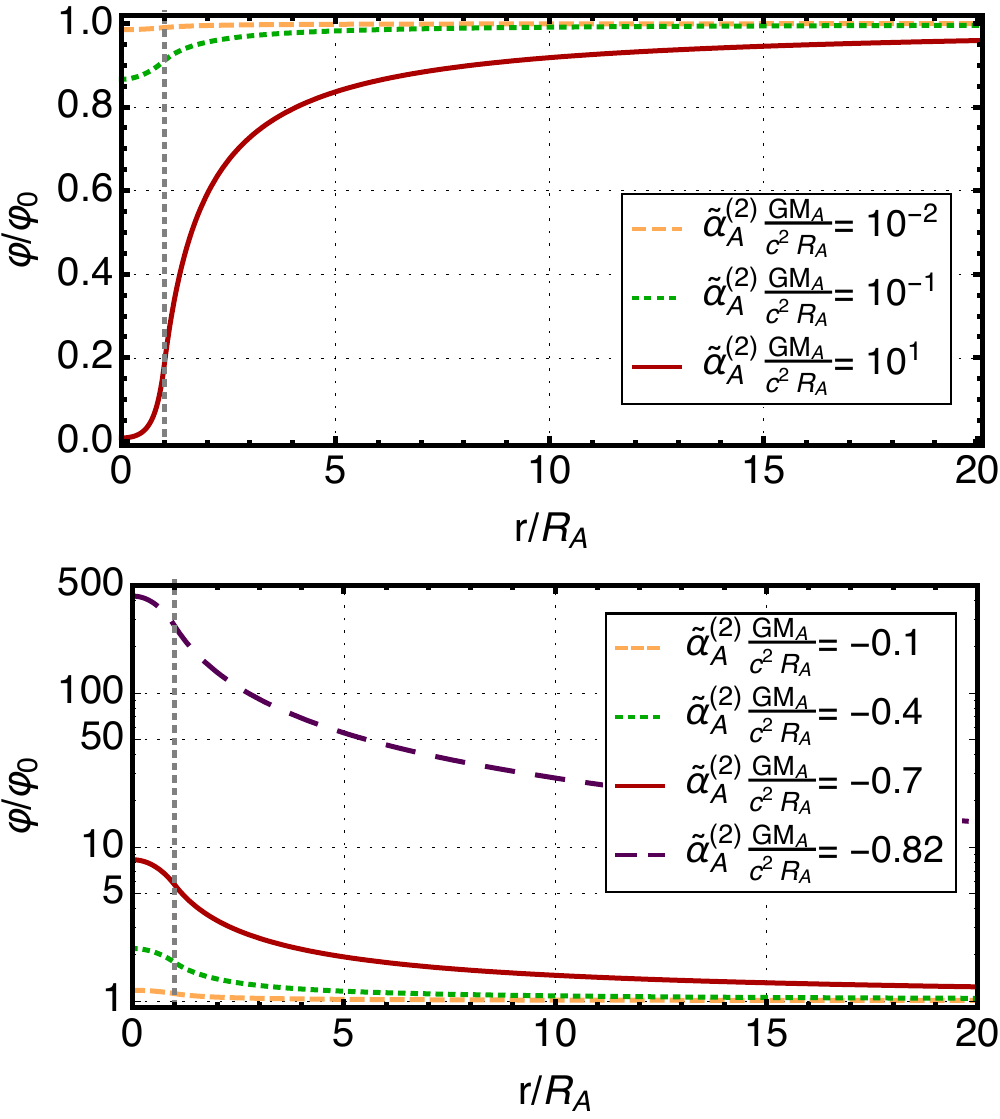}
\end{center}
\caption{Evolution of the scalar field around a homogeneous spherically symmetric body. The different curves show the impact of the values of $\tilde\alpha^{(2)}$. In particular, in the limit of large positive couplings, the scalar field tends to vanish inside the body and the scalar field diverges for negative values of $\tilde\alpha^{(2)}$.}
\label{fig:phi_quad}
\end{figure}

In particular, the sign of $\tilde \alpha_A^{(2)}$ (or of $\varepsilon_A$) plays an important role and two different nonlinear mechanisms can arise: a screening mechanism for $\varepsilon_A > 0$ and an amplification mechanism for $\varepsilon_A <  0$ (see Figs.~\ref{fig:scalar_charge} and \ref{fig:phi_quad}). This behavior is similar to that arising for massless scalar fields, for which both amplification and deamplification nonpertubative mechanisms have been studied since the seminal work of Damour and Esposito-Far\`ese~\cite{damour:1993vn,*damour:1996uq}. In particular, in metric theories, the amplification mechanism for $\tilde\alpha_A^{(2)} < 0$ has been known as the scalarization of compact objects.

For positive values of the coupling coefficient $\tilde\alpha_A^{(2)} >0$ and for very large couplings ($\varepsilon_A \gg 1$), one gets $J_+(x)\approx 3/x^2$. In that case, $s^{(2)}_A\approx  \frac{R_Ac^2}{GM_A}$ and the scalar field at the surface of the body ($r=R_A$ in Eq.~(\ref{eq:phi_2})) tends to vanish. Indeed, the scalar field solution in that limit reduces to
\begin{equation}
	\varphi^{(2)}(t,\bm x)=\varphi_0 \cos \left(\frac{m_\varphi c^2}{\hbar}t+\delta\right) \left( 1- \frac{R_A}{r}\right).
\end{equation}
Similarly, the interior solution tends to 0 when the coupling constant increases (see in Appendix~\ref{ap:scal_quad} for its expression, and see the top of Fig.~\ref{fig:phi_quad}). This means that the scalar field only penetrates a thin shell at the surface of the body. A detailed analysis of the interior solution given by Eq.~(\ref{eq:intern}) shows that the typical length over which the field is not constant inside the body is given by $\ell\sim R_A/ \left(3\tilde \alpha_A^{(2)}GM_A/c^2/R_A\right)^{1/2}$. Fig.~\ref{fig:phi_quad} illustrates this behavior, which has similarities with the chameleon mechanism \cite{khoury:2004uq,*khoury:2004fk,*hees:2012kx,*burrage:2018aa}. Conceptually, the situation can be compared to the case of an insulator located in an external electric field: the electric field inside and at the surface will vanish. This property has an interesting consequence: experiments located at the surface of the Earth are less suitable to detect or constrain such a scalar field in this regime while space-based experiments are better suited.  

On the other hand, for the cases where $\tilde\alpha_A^{(2)} <0$, the scalar field diverges in the limit where $\left|\tilde \alpha_A^{(2)}\right| \frac{GM_A}{c^2R_A} \rightarrow \frac{\pi^2}{12}$, as illustrated in the bottom of Fig.~\ref{fig:phi_quad} and in Fig.~\ref{fig:scalar_charge}. The Minkowskian approximation used to solve for the scalar field breaks down when $\varphi\sim 1$ (see the beginning of Sec.~\ref{sec:phi}). For couplings that lead to $\varphi >1$, one needs to self-consistently solve numerically all of the field equations, including the backreaction from the metric, in order to fully take into account nonlinear behavior. 

On top of that, when $d^{(2)}_i\varphi^2/2<-1$, the fundamental constants from Eq.~(\ref{eq:constants}) would change their sign, which would be an unacceptable behavior.  

The amplification mechanism for $\tilde\alpha_A^{(2)} < 0$ in metric theories has been known as the scalarization of compact objects \cite{damour:1993vn,*damour:1996uq}. It is a fully nonperturbative effect that requires us to solve for both the scalar and the metric field equations numerically. Recently, several works extended the work from \cite{damour:1993vn,*damour:1996uq} to the case of massive scalar fields \cite{chen:2015aa,ramazanoglu:2016pr,yazadjiev:2016rp,doneva:2016jc,morisaki:2017aa}. However, those studies only focus on stationary solutions of the field equations, preventing them from finding oscillating dark-matter candidate solutions to the problem. The solutions presented in this section, although only valid for weak gravitational fields, indicate that a nonstationary scalarization may also occur for light scalar DM. In other words, DM as a light scalar field may also lead to a potential scalarization of compact objects. A detailed investigation of such effects which would include the nonpertubative resolution of the scalar and the metric field equations without the stationarity assumption is beyond the scope of this paper.

\subsection{Identification as Dark Matter}
\label{sec:ID}
In order to identify the scalar field as DM, one has to consider its asymptotical behavior. For both solutions computed in the previous section, the scalar field oscillates at spatial infinity. It can be shown that this scalar field gives rise to  the following cosmological energy density $\rho_\varphi$ and pressure $p_\varphi$:
	\begin{align*}
		\rho_\varphi &=\frac{c^2}{8\pi G}\left[\dot\varphi^2+\frac{c^2V(\varphi)}{2}\right] \, , \\
		p_\varphi    &=\frac{c^2}{8\pi G}\left[\dot\varphi^2-\frac{c^2V(\varphi)}{2}\right] \, .
	\end{align*}
After averaging over one period, a coherently oscillating scalar field gives a vanishing pressure and an energy density \cite{van-tilburg:2015fj,hees:2016uq}
\begin{equation}\label{eq:rhoDM}
	\rho_\varphi=\frac{c^6}{4\pi G \hbar^2}\frac{m^2_\varphi \varphi_0^2}{2}\, .
\end{equation}
Assuming that all DM is made of one light scalar field, this relationship fixes its amplitude for a given mass. Using a value for the local galactic DM energy density of $\rho_\varphi= 0.4$ GeV/cm$^3$ \cite{mcmillan:2011vn}, one gets that the amplitude of the scalar field oscillation at infinity is given by \cite{van-tilburg:2015fj,hees:2016uq}
\begin{equation}
\varphi_0\sim \frac{7\times 10^{-31} \textrm{ eV}}{m_\varphi}\, .
\end{equation}
Considering that cosmological observations put a lower limit on the scalar field mass at the level of $10^{-24}-10^{-22}$ eV (assuming that these scalar fields saturate the observed DM content) \cite{hu:2000aa,beyer:2014aa,urena-lopez:2016aa,calabrese:2016aa,marsh:2016aa,bernal:2017aa,hui:2017aa,bernal:2018aa}, $\varphi_0$ is always smaller than $7 \times 10^{-7}$, justifying the Minkowskian approximation used in this section.


\section{Observables}
\label{sec:obs}
\subsection{Comparison of two atomic clocks}
One way to search for a violation of the EEP is to measure the frequency ratio between two clocks working on different atomic transitions and located at the same position. The observable is then $Y=X_A/X_B$ where $X_A$ and $X_B$ are the specific transitions for each clock. It follows from Eq.~(\ref{eq:clock_def}) that the relative variation of $Y$  ($Y/Y_0$) takes the form of
\begin{equation}
	d \ln \frac{Y}{Y_0} = \left(\kappa^{(i)}_{X_A} - \kappa^{(i)}_{X_B}\right)d \left(\varphi^i\right)\, .
\end{equation} 
If we assume that the variations are small (i.e. $\left|Y/Y_0-1\right| \ll 1 $), then the evolution of the observable is given by
\begin{equation}
	 \frac{Y(t,\bm x)}{Y_0} = K + \left(\kappa^{(i)}_{X_A} - \kappa^{(i)}_{X_B}\right) \varphi^i(t,\bm x) \, ,
\end{equation}
where $K$ is a constant that is unobservable.

\subsubsection{Linear coupling}
Using the expression of the scalar field solution of the Klein-Gordon equation with a linear coupling from Eq.~(\ref{eq:phi_1}) leads to
\begin{align}
	\frac{Y(t,\bm x)}{Y_0} &= K + \Delta \kappa^{(1)} \varphi_0 \cos \left(\bm k .\bm x -\omega t+\delta\right) \label{eq:clock_lin} \\
	&\quad  - \Delta \kappa^{(1)} s^{(1)}_A \frac{GM_A}{c^2r}e^{-r/\lambda_\varphi} \, .\nonumber
\end{align}
The first part corresponds to the coupling of the clocks to the oscillating DM field. This signature has already been searched for in several measurements \cite{van-tilburg:2015fj,hees:2016uq,kalaydzhyan:2017aa}. The second part corresponds to the coupling of the clock to the scalar field generated by the central body and has been considered in data analysis in \cite{leefer:2016aa}.

\subsubsection{Quadratic coupling}
The signature produced by the scalar field in the case of a quadratic coupling between the scalar field and matter is richer. Using the scalar field solution from Eq.~(\ref{eq:phi_2}), it reads
\begin{align}
	\frac{Y(t,\bm x)}{Y_0} &= K + \Delta \kappa^{(2)}\frac{\varphi_0^2}{2}\left(1-s^{(2)}_A\frac{GM_A}{c^2r}\right)^2 \label{eq:clock_quad}\\ 
	& +\Delta \kappa^{(2)}\frac{\varphi_0^2}{2} \cos \left(2\omega t +2\delta\right)\left(1-s^{(2)}_A\frac{GM_A}{c^2r}\right)^2\,\nonumber
\end{align}
where $\omega=m_\varphi c^2/\hbar$. This signature is quite unique and is the combination of two distinct terms. The first one is space dependent and could be searched for by comparing spatial and terrestrial clocks located at various positions, or by monitoring the evolution of the frequency of a given clock orbiting in an eccentric orbit around Earth. The second term is an oscillating term whose amplitude depends on the location in the gravitational field as well.
In particular, if one considers two clocks located at the surface of the Earth ($r=R_\oplus$), the oscillating part of the signal becomes (from Eqs. (\ref{eqapp:varphiansatz}), (\ref{eqapp:solX}) and (\ref{eqapp:Kp}))
\begin{align}
	\tilde Y(t)&=\frac{\Delta \kappa^{(2)}}{\tilde \alpha^{(2)}_\oplus}\frac{c^2R_\oplus}{6 GM_\oplus} \varphi_0^2 \tanh^2\left(\sqrt{3\tilde \alpha_\oplus^{(2)} \frac{GM_\oplus}{c^2R_\oplus}}\right) \label{eq:Yosc} \\
	&\qquad \qquad \times \cos \left(2\omega t +2\delta\right) \,  \nonumber
\end{align}
for a positive $\tilde\alpha_\oplus^{(2)}$.

\subsection{Tests of the Universality of Free Fall}
The motion of a test mass can be derived from the action (\ref{eq:smat}), or equivalently from the Lagrangian
\begin{equation}
	L_T = -m_T(\varphi)c\sqrt{-g_{\mu\nu}\frac{dx^\mu}{dt} \frac{dx^\nu}{dt}}\, .
\end{equation}
In this section, we are interested in UFF experiments for which the acceleration of two test masses located at the same location are compared. Therefore, we are only interested in the first-order part of the acceleration that is composition dependent. We can therefore use the following approximation for the Lagrangian:
\begin{equation}
	L_T\approx-m_T(\varphi)c^2\left(1-\frac{v^2}{2c^2}\right) \, ,
\end{equation}
where, as for the Klein-Gordon equation, we neglect terms that are of the order of $\mathcal O\left(\left| h_{\mu\nu}\right|\right)$. A simple Euler-Lagrange derivation gives the first-order contribution to the violation of the UFF:
\begin{equation}
	\left[ \bm a_T \right]_\textrm{EEP}= -\alpha_T(\varphi)\left[c^2 \bm \nabla \varphi + \bm v \dot \varphi \right]\, ,
\end{equation}
where $\alpha_T$ is the coupling defined by Eq.~(\ref{eq:alpha}). The differential acceleration between two bodies $A$ and $B$ located at the same position is therefore given by
\begin{align}
	\left[\Delta \bm a\right]_{A-B}&=\bm a_A(t,\bm x)-\bm a_B(t,\bm x) \nonumber \\
	& = -\left(\alpha_A(\varphi)-\alpha_B(\varphi)\right)\left[c^2 \bm \nabla \varphi + \bm v \dot \varphi \right]\,. 
\end{align}

\subsubsection{Linear coupling}
In the case of a linear coupling, the differential acceleration between two bodies $A$ and $B$ located in the same location, in the gravitational field generated by a central body $C$ can be determined from Eq.~(\ref{eq:phi_1}) and is given by
\begin{align}
	\left[\Delta \bm a\right]_{A-B}&=\Delta \bar \alpha^{(1)}\varphi_0\left( c^2 \bm k-\omega \bm v \right)\sin \left(\bm k.\bm x -\omega t +\delta\right)  \nonumber\\
	& - \Delta \bar \alpha^{(1)} s_C^{(1)} e^{-r/\lambda_\varphi}\left(1+\frac{r}{\lambda_\varphi}\right)\frac{GM_C}{r^3} \bm x \, ,\label{eq:UFF_lin}
\end{align}
where $\Delta \bar \alpha^{(1)}=\left(\bar \alpha^{(1)}_A -\bar \alpha^{(1)}_B\right)$ with $\bar \alpha^{(1)}$ given in Eq.~(\ref{eq:bar_alpha}) and where $s_C^{(1)}$ is defined in Eq.~(\ref{eq:s1}) and depends linearly on $\bar\alpha^{(1)}_C$. The first line represents an oscillating variation of the differential acceleration of the two bodies. This oscillation is induced by the oscillating DM. The amplitude of this UFF violation is linearly proportional to the coupling constant $d^{(1)}_i$. The second line is a regular fifth-force differential acceleration that is due to the coupling of the two bodies to the scalar field generated by the central body. The amplitude of the violation of the UFF is proportional to the square of the $d_i^{(1)}$ coefficients. This term can be identified from standard UFF measurements by using the E\"otv\"os parameter $\eta$ defined as 
\begin{equation}\label{eq:eta}
	\eta=2 \frac{\left| \bm a_A- \bm a_B\right|}{\left| \bm a_A+ \bm a_B\right|} \, ,
\end{equation}
by (see also \cite{wagner:2012fk})
\begin{equation}\label{eq:eta_lin}
	\eta = \Delta \tilde \alpha^{(1)} s_C^{(1)} e^{-r/\lambda_\varphi}\left(1+\frac{r}{\lambda_\varphi}\right)\, .
\end{equation}

\subsubsection{Quadratic coupling}
The differential acceleration between two bodies in the case of a quadratic coupling can be determined from Eq.~(\ref{eq:phi_2}) and is given by
\begin{align}
	\left[\Delta \bm a\right]_{A-B}&=\Delta \bar \alpha^{(2)}\frac{\varphi_0^2}{2} \left(1-s_C^{(2)}\frac{GM_c}{c^2r}\right)\Bigg[ -\frac{GM_c}{r^3}\bm x \  s_C^{(2)}  \nonumber\\
	& -\frac{GM_c}{r^3}\bm x  s_C^{(2)}  \cos\left(2\omega t+2\delta\right) \label{eq:UFF_quad} \\
	& + \left(1-s_C^{(2)}\frac{GM_c}{c^2r}\right)\omega\bm v\sin \left(2\omega t+2\delta\right)\Bigg]\nonumber
\end{align}
where $\Delta \bar \alpha^{(2)}=\left(\bar \alpha^{(2)}_A -\bar \alpha^{(2)}_B\right)$ with $\bar \alpha^{(2)}$ given in Eq.~(\ref{eq:bar_alpha}) and $s^{(2)}_C$ defined in Eq.~(\ref{eq:sa}). The first line corresponds to a  differential acceleration proportional to the Newtonian acceleration and arises from the gradient of the DM field density induced by the central body. This term can be identified from standard UFF measurements by using the E\"otv\"os parameter $\eta$ defined by Eq.~(\ref{eq:eta}) with
\begin{equation}
	\eta= s_C^{(2)} \ \Delta \bar \alpha^{(2)}\frac{\varphi_0^2}{2} \left(1-s_C^{(2)}\frac{GM_c}{c^2r}\right)\, .\label{eq:eta2}
\end{equation}
Note that in the small coupling case and/or in remote regions with respect to the source, the E\"otv\"os parameter reduces to
\begin{equation}
	\eta \approx s_C^{(2)} \ \Delta \bar \alpha^{(2)}\frac{\varphi_0^2}{2}.
\end{equation}
Hence, in these regimes, the E\"otv\"os parameter becomes independent of the location of the two test masses with respect to the source of the gravitational field. This corresponds to the standard parametrization used for tests of the UFF (see e.g. \cite{will:2014la}).

On the other hand, in the neighborhood region of a central body and in the limit of strong couplings, the E\"otv\"os parameter grows linearly with the altitude $h$ (with respect to the radius $R_A$). Indeed, in the strong coupling case (see Sec.~\ref{sec:sol_quad}), the E\"otv\"os parameter can be rewritten in terms of the altitude $h$ as follows
\begin{equation}
	\eta \approx s_C^{(2)} \ \Delta \bar \alpha^{(2)}\frac{\varphi_0^2}{2} \frac{h}{R_A+h}.
\end{equation}
This is another unique feature that could potentially  be tested. But since the E\"otv\"os parameter becomes independent of the location for $h \gg R_A$, designing an experiment in this regime would be especially constraining for theories with strong couplings, whereas they are much less constrained in regimes such that $h \ll R_A$ -- such as in MICROSCOPE-like configurations for instance.

The second and third lines of Eq.~(\ref{eq:UFF_quad}) correspond to an oscillating differential acceleration whose amplitude depends on the radial coordinates. The amplitude of the oscillation in the differential oscillation from the second line is the same as the static violation of the UFF from the first line. These oscillations could be searched for in UFF measurements as discussed in Sec.~\ref{sec:const_quad}.

\section{Constraints from existing measurements}
\label{sec:constraints}
In this work, we present the results using the method of ``maximum reach analysis'' (MRA), which consists of varying the parameters one at a time, while the others are kept equal to zero. 
This method allows us to obtain an idealistic estimate of the parameters' limit. It has also the benefit of producing readable 2D plots of the constraints. More information about this method that is implicitly used in most papers using atomic sensors \cite{stadnik:2015aa,stadnik:2015yu,arvanitaki:2016qv}, can be found in \cite{bourgoin:2016yu,*bourgoin:2017aa}. 

When the amplitude of the scalar field oscillation $\varphi_0$ is needed, we assume that the scalar field comprises all the DM, which has a local energy density of $\rho_\varphi=0.4$ GeV/cm$^3$ \cite{mcmillan:2011vn}. This value fixes $\varphi_0$ through Eq.~(\ref{eq:rhoDM}). Similarly, the velocity of DM used in our calculation is $\left|\bm v\right|=10^{-3}$ c.

\subsection{Description of the existing measurements}
We will now compare the signatures described in the previous section to existing measurements. In this paper, we will use four different types of measurements:
\begin{itemize}
	\item Measurements of the UFF done by the E\"ot-Wash laboratory~\cite{schlamminger:2008zr,wagner:2012fk}. This consists of a measurement of the differential acceleration between two test masses at the Earth's surface. Two types of pairs of test masses have been used: (i) Be versus Ti and (ii) Be versus Al. The dilatonic charges related to these elements are given in Tab.~\ref{tab:dilaton_coef}, as well as the dilatonic charge of the Earth, which is assumed to be composed of an iron core and a silicate mantle \cite{damour:2010zr}. For each of these pairs of test masses, a violation of the UFF in the field of the Earth and in the field of the Sun has been searched for. This provides 4 different constraints in total, which are summarized in Tab.~\ref{tab:UFF}.
	
	\item To probe the UFF at very short distances, the E\"ot-Wash group also performed an experiment where they made a body of Uranium rotate around the test masses to search for a violation of the UFF in the gravitational field of that body \cite{smith:1999aa}. This measurement also has the advantage of being sensitive to different linear combinations of the matter-scalar coupling coefficients. For this particular experiment, the test masses are made of Cu and Pb and the $^{238}$U source is located 10.2 cm from the test masses. The constraint obtained on the $\eta$ parameter is mentioned in Tab.~\ref{tab:UFF}.
	
	\item The first reported measurements of the UFF performed by the MICROSCOPE space mission~\cite{touboul:2017aa}. This consists of a measurement of the differential acceleration between two test masses orbiting around the Earth on a nearly circular orbit (710 km of altitude, where the Earth's gravitational acceleration is 7.9 m/s$^{2}$). The two test masses are made of an alloy of Pt and Ti. The exact composition and the related dilatonic charges for these test masses are given in Tab.~\ref{tab:dilaton_coef}. The first result from MICROSCOPE is given in Tab.~\ref{tab:UFF}.

	\item We search for oscillatory signatures in the comparison between two frequencies delivered by two different atomic transitions. This kind of measurement compares directly the frequencies delivered by two different hyperfine or radio-frequency transitions at the same location in space. Such measurements using two isotopes of Dysprosium and the related data analysis have been performed in~\cite{van-tilburg:2015fj}. Measurements of the the dual Rubidium-Cesium atomic fountain from SYRTE (for a description of the experiment, see~\cite{guena:2010kx,guena:2012vn,guena:2014fr}) have also been used to search for such an oscillation~\cite{hees:2016uq}. In these measurements, the raw data consist of a frequency comparison $Y$ between two different transitions and the analysis from~\cite{van-tilburg:2015fj,hees:2016uq} consists of fitting a harmonic model $\mathcal A_\omega \cos (\omega t+\delta)$ to the data for different frequencies (limited by the the data span and the Nyquist frequency, although a method has been suggested to search for periodic variations beyond that limit, see~\cite{kalaydzhyan:2017aa}). An upper limit on the amplitude $\mathcal A_\omega$ as a function of $\omega$ is the result of these analyses (see \cite{van-tilburg:2015fj,hees:2016uq}).

\end{itemize} 
\begin{table}[htb]
 \caption{Measurement of the E\"otv\"os parameter from the E\"ot-Wash laboratory (see~\cite{schlamminger:2008zr,wagner:2012fk}) and from MICROSCOPE~\cite{touboul:2017aa}. $\eta_{A-B;C}$ refers to Eq.~(\ref{eq:eta}) and quantifies the differential acceleration between two bodies $A$ and $B$ in the gravitational field generated by the body $C$. The given uncertainties correspond to the 1-$\sigma$ uncertainties.}
	\label{tab:UFF} 
	\centering
	\begin{tabular}{c c c }
	\hline
	     & Measurements  & Reference  \\
	\hline
	 $\eta_{\textrm{Be-Ti;}\oplus}$ & $(0.3\pm 1.8)  \times 10^{-13}$ & \cite{schlamminger:2008zr,wagner:2012fk} \\
	 $\eta_{\textrm{Be-Ti;}\odot}$  & $(-3.1\pm 4.7) \times 10^{-13}$ & \cite{wagner:2012fk} \\
	 $\eta_{\textrm{Be-Al;}\oplus}$ & $(-0.7\pm 1.3)  \times 10^{-13}$ & \cite{wagner:2012fk} \\
	 $\eta_{\textrm{Be-Al;}\odot}$  & $(-5.2\pm 4.0) \times 10^{-13}$ & \cite{wagner:2012fk} \\ \hline
	 $\eta_{\textrm{Cu-Pt;}\ ^{238}\textrm{U}}$  & $(1.1\pm 3.0) \times 10^{-9\phantom{1}}$ & \cite{smith:1999aa} \\ \hline
	 $\eta_{\textrm{Pt-Ti;}\oplus}$ & $(-0.1\pm 1.3) \times 10^{-14}$ & \cite{touboul:2017aa} \\ 
	\hline
	\end{tabular}
\end{table}

Although current measurements of the UFF using atom interferometry \cite{fray:2004fk,*merlet:2010ys,*bonnin:2013aa,*schlippert:2014yq,*tarallo:2014fj,*zhou:2015vn} are not as constraining as macroscopic measurements, with future improvements, these can also be used to search for ultra-light DM. In particular, microscopic UFF measurements performed in space, like proposed in e.g. STE-QUEST \cite{altschul:2015sh}, will be very adapted to search for such a DM candidate.

\subsection{Linear coupling}
\label{sec:obs_lin}
\begin{figure}[hbt]
\begin{center}
\includegraphics[width=0.45\textwidth]{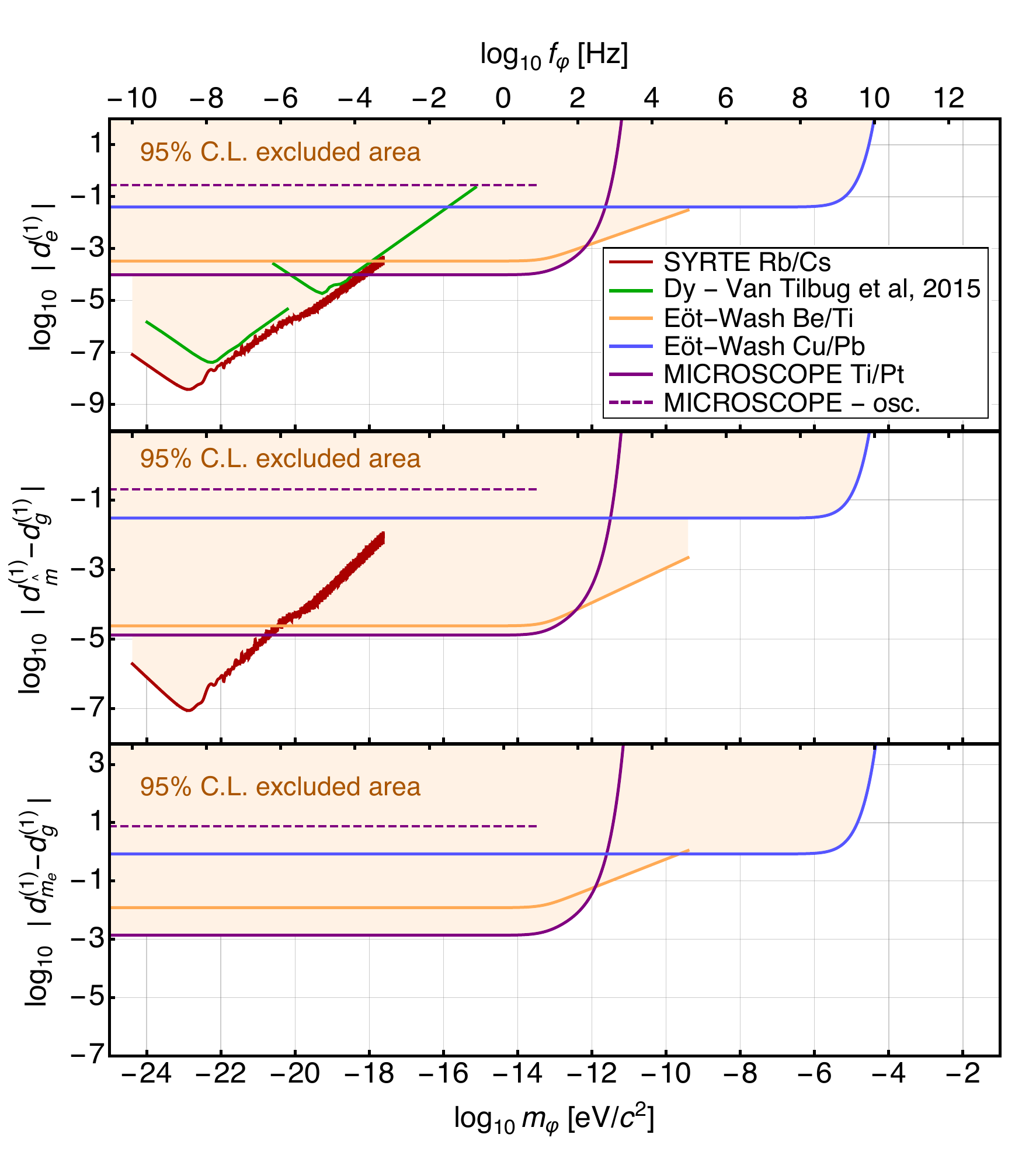}
\end{center}
\caption{Upper (MRA) limit (at 95\% confidence level) on the various scalar/matter coupling coefficients in the case of a linear coupling between matter and the scalar field. The SYRTE Cs/Rb analysis is from \cite{hees:2016uq}, the Dy analysis is presented in \cite{van-tilburg:2015fj}, the UFF measurement around Earth between Be and Ti is from \cite{schlamminger:2008zr}, the UFF measurement between Cu and Pb in the gravitational field of a $^{238}$U body is from \cite{smith:1999aa}, MICROSCOPE's result is presented in \cite{touboul:2017aa,berge:2018aa}. The constraints derived from clock measurements assumed that the scalar field comprises all local DM while the UFF constraints do not rely on this assumption. Note that the dashed line is not an actual constraint but an  estimate of the potential sensitivity, see Sec.~\ref{sec:obs_lin}.}
\label{fig:lin}
\end{figure}

First of all, the measurements of the ``static''  UFF which are summarized in Tab.~\ref{tab:UFF} can be directly used to constrain the E\"otv\"os parameter whose expression is given by Eq.~(\ref{eq:eta_lin}). This type of measurement will actually constraint combinations of the product of two constants $d_j^{(i)}$, in particular, $D^{(1)}_{\hat m}=d^{*(1)}_g\left(d^{(1)}_{\hat m}-d^{(1)}_g\right)$, $D^{(1)}_e=d^{*(1)}_g d_e$, $D^{(1)}_{m_e}=d^{*(1)}_g\left(d^{(1)}_{m_e}-d^{(1)}_g\right)$, etc. Combined constraints on those variables have been presented in \cite{wagner:2012fk} for the E\"otwash measurements and in \cite{berge:2018aa} for the MICROSCOPE results. The MRA analysis using the UFF measurements are presented in Fig.~\ref{fig:lin}. They are constant for small masses, up to $\lambda_\varphi$ of the order of the Earth radius. For small masses, the interaction length $\lambda_\varphi$ is larger than the distance between the test masses and the center of the Earth. For these distances, the Earth can be considered as a point mass and the measurements from the MICROSCOPE satellite are the most constraining. For large masses, the interaction length $\lambda_\varphi$ is smaller than the distance between the test masses and the center of the Earth, and experiments at the Earth's surface become more sensitive. Nevertheless, both decrease when $\lambda_\varphi$ decreases (or equivalently when $m_\varphi$ increases), but with different initial slopes (in a log-log scale) that are directly related to the distance between the two test masses and the center of the Earth. The slope is initially more favorable for experiments at the surface of the Earth. For short distances ($\lambda\sim 10$ km), the Yukawa field will be determined by the local environment and by the topography around the experiment. Therefore, UFF constraints in the field of the Earth are limited to large values of $\lambda$ in Fig.~\ref{fig:lin}. For very short distances, the dedicated experiment measuring the differential acceleration in the gravitational field generated in the lab by a uranium body \cite{smith:1999aa} is the most powerful, see Fig.~\ref{fig:lin}.

Note that, contrary to clock measurements, the UFF constraints are independent of the hypothesis that the scalar field discussed here constitutes the DM in our Galaxy.

The results from atomic clock measurements, also presented in Fig.~\ref{fig:lin}, are fully detailed in \cite{van-tilburg:2015fj,hees:2016uq}. They are obtained by equating the amplitude of the oscillation from Eq.~(\ref{eq:clock_lin}) to the upper limit of the amplitude $\mathcal A_\omega$ fitted to the data. For small masses, they are more constraining than UFF measurements. For masses larger than $10^{-22}$ eV, the upper limits on the $d_i$ increase linearly with the mass of the scalar field. It is worth mentioning that these constraints depend on the identification of the scalar field as the unique component of DM (because it fixes the amplitude $\varphi_0$ through Eq.~(\ref{eq:rhoDM})), while the UFF constraints from the previous paragraph are obtained independently of the DM interpretation of the scalar field. Note also that constraints from the considered clock measurements are practically insensitive to $d_{m_e}^{(1)}-d_g^{(1)}$.

In addition to searching for an oscillation with atomic sensors, one can search for a Yukawa dependency of the comparisons between frequencies (see the second line of Eq.~(\ref{eq:clock_lin})), when clocks are moved in a given gravitational field.

Such a scenario has been considered in \cite{leefer:2016aa} but is currently not as competitive as the other measurements for individual coupling parameters.

Note that the so-called \textit{natural} couplings -- usually defined as couplings of the order of unity -- are excluded for scalar field masses $m_\varphi$ up to  $\sim 10^{-5}$ eV for $d_e^{(1)}$, up to $\sim 10^{-4}$ eV for $d_{\hat m}^{(1)}-d_g^{(1)}$ and up to $10^{-5}$ eV for $d_{m_e}^{(1)}-d_g^{(1)}$.

Finally, it is interesting to mention that UFF measurements could potentially be reanalyzed to search for a periodic variation in the signal that would come from the first term of Eq.~(\ref{eq:UFF_lin}) \cite{graham:2016ab}. The dashed line from Fig.~\ref{fig:lin} represents an estimate of the sensitivity that could be obtained on the various coefficients if the amplitude of the UFF oscillation from Eq.~(\ref{eq:UFF_lin}) could be constrained at the level of $\delta a \sim \eta g$ where $g$ is the gravitational acceleration. In Fig.~\ref{fig:lin}, we present the constraints that can be obtained from a reanalysis of MICROSCOPE observations in the frequency range that seems reachable from the current measurements. However, one can see from Fig.~\ref{fig:lin} that they would not be competitive to existing constraints.

\subsection{Quadratic coupling}\label{sec:const_quad}
The case of quadratic coupling is more complex and offers a richer phenomenology. The reason for this richer phenomenology comes from the presence of the $J_\pm(x)$ factor in the expression of the effective scalar charge $s^{(2)}_A$ in Eq.~(\ref{eq:sa}). First of all, as discussed in Sec.~\ref{sec:sol_quad}, the behavior of the effective scalar charge is significantly different for positive and negative couplings (this is illustrated in Fig.~\ref{fig:scalar_charge}). 

\begin{figure*}[htb]
\begin{center}
\includegraphics[width=0.45\textwidth]{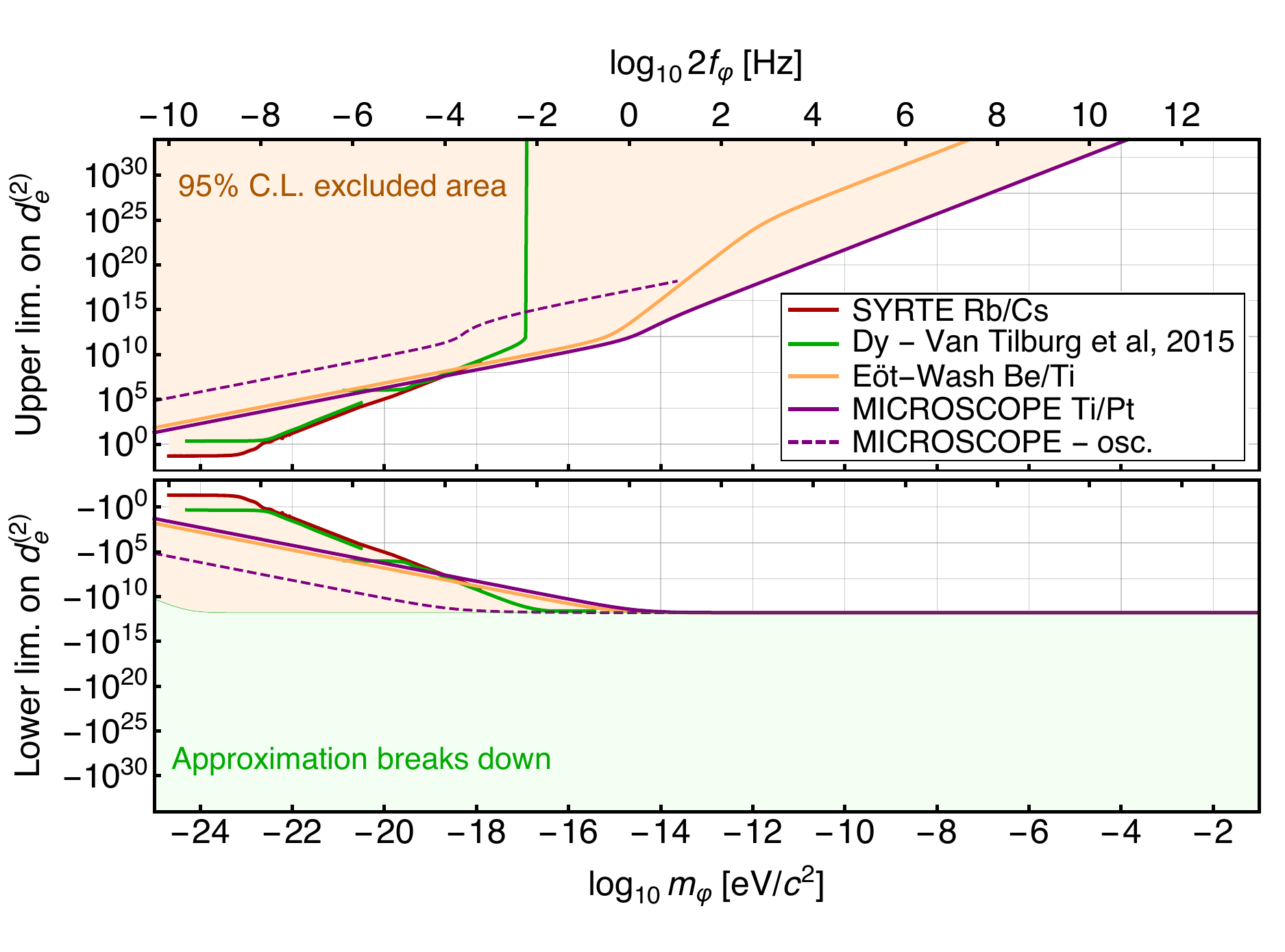}\hfill
\includegraphics[width=0.45\textwidth]{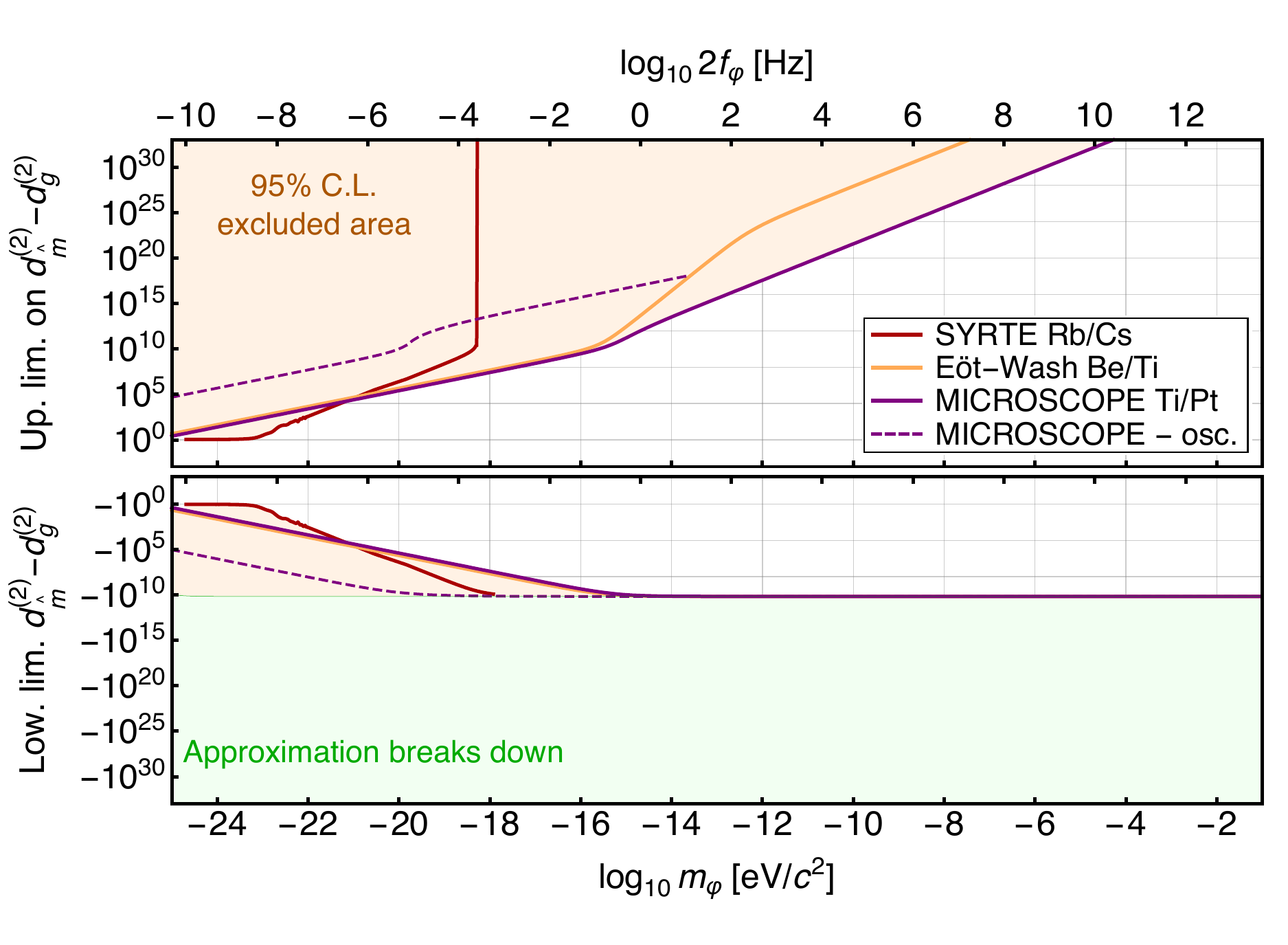}
\includegraphics[width=0.45\textwidth]{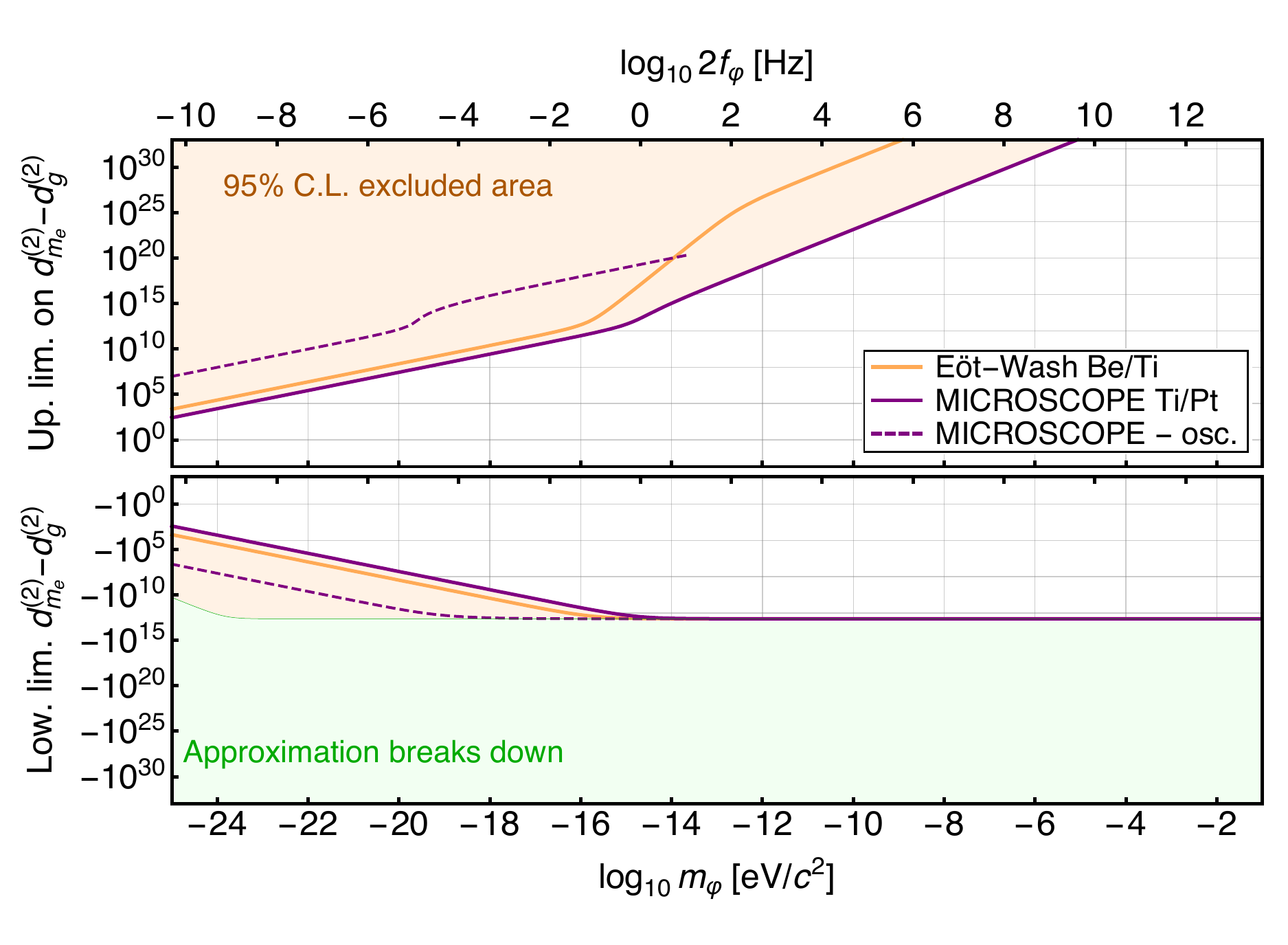}
\end{center}
\caption{Upper and lower (MRA) limits (at 95\% confidence level) on the various scalar/matter coupling coefficients $d_i^{(2)}$ in the case of a quadratic coupling between matter and the scalar field. The constraints have been derived using the following measurements: the SYRTE Cs/Rb data from \cite{hees:2016uq}, the Dy measurements from \cite{van-tilburg:2015fj}, the UFF measurement around Earth between Be and Ti from \cite{schlamminger:2008zr} and the MICROSCOPE's result presented in \cite{touboul:2017aa}. Note that the dashed line is not an actual constraint but an estimate of the potential sensitivity that would be obtained by searching for an oscillating violation of the UFF within MICROSCOPE data. The lower green shaded area represents the limits for which $\left|d_i\varphi^2/2\right|\sim 1$ where the Minkowskian approximation used in this work breaks down and where the constants of Nature from Eq.~(\ref{eq:constants}) would naively experience a change of sign.}
\label{fig:quad}
\end{figure*}

In particular, for large positive couplings $d_i^{(2)}$, a screening mechanism occurs: the amplitude of the oscillations of the scalar field at the surface of the central body decreases, lowering the sensitivity of any measurement. This kind of effect occurs when $\varepsilon_A \sim d_i \left[Q_i\right]_A \frac{GM_A}{c^2R_A}\sim 1$, which corresponds to the case where the $s^{(2)}_A \frac{GM_A}{c^2r}$ part of the scalar field solution from Eq.~(\ref{eq:phi_2}) starts to become relevant. This deamplification mechanism discussed in Sec.~\ref{sec:sol_quad} makes the scalar field hard to detect and constrain for large scalar field masses as can be seen in Fig.~\ref{fig:quad}. 

On the other hand, the case of large negative couplings $d_i^{(2)}<0$ is characterized by an amplification of the scalar field (see Figs.~\ref{fig:scalar_charge} and \ref{fig:phi_quad}) that increases the amplitude of observables, which makes DM easier to either detect or constrain. As mentioned in Sec.~\ref{sec:sol_quad}, at some point when the scalar field becomes too large, the approximation used in this work breaks down. In particular, the development done in \cite{damour:2010ve,damour:2010zr} requires that $d_i\varphi^2/2 < 1$ so that the variation of the constants of Nature from Eq.~(\ref{eq:constants}) can be treated perturbatively. Moreover, the limit  $d_i\varphi^2/2=-1$ would naively imply a change of the sign of the constants of Nature, an undesirable behavior. This limit -- where the approximation used in this work breaks down, and where the constants of Nature would change their sign -- is indicated in Fig.~\ref{fig:quad} by a shaded green area. A full understanding of the behavior for large negative couplings requires us to extend the work of \cite{damour:2010ve,damour:2010zr} at the nonperturbative level and to solve the full relativistic field equations nonperturbatively (as discussed in Sec.~\ref{sec:sol_quad}).

In Fig.~\ref{fig:quad}, clock measurements from \cite{van-tilburg:2015fj,hees:2016uq} have been transformed into constraints on the $d_i^{(2)}$ coefficients. In order to do so, the published constraints on $d_i^{(1)}$ need to be transformed into a constraint on the  amplitude $\mathcal A_\omega$ of an oscillation that has been constrained from the data by using the first line of Eq.~(\ref{eq:clock_lin}), and then transformed back into a constraint on the $d_i^{(2)}$ coefficients by using the second line of Eq.~(\ref{eq:clock_quad}). For small scalar field masses $m_\varphi$, the upper and lower limits on $d_i^{(2)}$ coefficients evolve quadratically with $m_\varphi$ and agree with those derived in \cite{stadnik:2015aa,stadnik:2016kq}. At some point, when the constraints reach a value of $d_i^{(2)}$ that is such that $\varepsilon_A=d_i^{(2)}\left[Q_i\right]_A\frac{GM_A}{c^2R_A}\sim 1$, the $s^{(2)}_A \frac{GM_A}{c^2r}$ part of the scalar field solution from Eq.~(\ref{eq:phi_2}) starts to become relevant and the behavior of the constraint becomes dependent on the sign of $d^{(2)}_i$. For negative $d^{(2)}_i$, the scalar field and the observables diverge, which produces a saturation of the constraints. For large positive $d^{(2)}_i$, it can be seen from Eq.~(\ref{eq:Yosc}) that the amplitude of the oscillation on the clock observable evolves as $\propto 1/m_\varphi^2$ and becomes independent of $d_i^{(2)}$ -- because  $\Delta \kappa^{(2)}/\tilde \alpha_\oplus^{(2)}$ is constant (see Eqs.~(\ref{eq:tilde_alpha})-(\ref{eq:bar_alpha}) and (\ref{eq:kappa})) and $\varphi_0 m_\varphi$ is as well (see  Eq.~(\ref{eq:rhoDM})). In that particular limit, there exists a critical scalar mass above which the amplitude of the scalar field-induced oscillation on clock measurements becomes smaller than the limit measured. Therefore, above this critical mass, no constraint on the $d_i^{(2)}$ parameters can be inferred, as can be seen on Fig.~\ref{fig:quad}. For small masses, clock results are the most constraining measurements.  Similarly to the linear case, the considered clock measurements are not sensitive to $d_{m_e}^{(2)}-d_g^{(2)}$.

Regarding the measurements of UFF violations, they can be separated into two parts: a static part and an oscillatory part. The UFF static measurements of $\eta$ from \cite{schlamminger:2008zr,touboul:2017aa} can directly be used and compared to the static expression of $\eta$ from Eq.~(\ref{eq:eta2}). It is important to notice that, contrary to the constraint on the $d^{(1)}_i$ coupling coefficients discussed in the previous section, the constraints on the $d_i^{(2)}$ parameters deduced from $\eta$ now depend on $\varphi_0$ and on the assumption that the scalar field is the unique component of DM. For small scalar field masses $m_\varphi$, the upper and lower limits on $d_i^{(2)}$ coefficients evolve linearly with $m_\varphi$. Similarly to the clock results, when $\varepsilon_A=d_i^{(2)}\left[Q_i\right]_A\frac{GM_A}{c^2R_A}\sim 1$ the constraints become sign dependent. For negative $d_i$, the scalar field and the observable diverge, which produces a saturation of the constraints. For large positive $d_i$, constraints evolve as $m_\varphi^2$ when taking into account the elevation of the measurements (see Fig.~\ref{fig:quad}). The reason why measurements at the surface of the central body are less constraining is related to the deamplification mechanism discussed in Sec.~\ref{sec:sol_quad}: the central body will act as an insulator for the scalar field and strongly reduces the scalar field at its surface (see Fig.~\ref{fig:phi_quad}), making it more difficult to detect. This is a major difference with linear couplings, for which measurements on Earth are more constraining for large scalar field masses. The measurements in space from MICROSCOPE are therefore the most constraining for quadratic couplings on a large mass range, often by several orders of magnitude.

It is worth mentioning that, contrary to the linear coupling case, values corresponding to so-called ``natural couplings'' (i.e. $d_i^{(2)}$ of the order of unity) are either not constrained at all, or only very marginally constrained for extremely small DM masses. This leaves a lot of space for so-called ``natural'' models to exist in the context of quadratic couplings.

Finally, it is also possible to reanalyze UFF experiments to search for harmonic oscillations in the data. Two types of signature can be searched for. The first one is related to the middle line of Eq.~(\ref{eq:UFF_quad}).  If one assumes that the limit on the amplitude of oscillations on $\delta a$ that can be reached using the MICROSCOPE data is $\sim \eta g$, one finds that searching the MICROSCOPE observations for such harmonic signatures could potentially produce other constraints on the $d_i^{(2)}$ coefficients at a similar level to the ones from the static UFF case (solid purple line in Fig.~\ref{fig:quad}). In addition to that, a second harmonic signature is produced by the third line of Eq.~(\ref{eq:UFF_quad}). Under the same assumption as above, if the MICROSCOPE observations are reanalyzed, they could produce constraints on the $d_i^{(2)}$ coefficients that are given by the dashed purple line in Fig.~\ref{fig:quad}. This sensitivity is nevertheless several orders of magnitude worse than that already existing from the static UFF analysis.

\section{Conclusion}
In this paper, we have studied the observable consequences induced by a violation of the Einstein Equivalence Principle for models of ultralight scalar DM in detail. We focused on two cases: (i)  a linear interaction between the DM scalar field and the standard model fields and (ii) a quadratic coupling between the scalar field and the standard model fields. The microscopic interactions between the scalar field and matter are modeled as in \cite{damour:2010ve,damour:2010zr}.

The specificity of our work is that we consider a massive scalar field, such that it can be identified as DM, and explore all local phenomenological consequences of such a field. We assume a mass range between $10^{-24}$ eV and $\sim$ eV, where the field would behave classically, and oscillate because of its potential. We show that, in particular for the quadratic coupling, this leads to new and unexpected phenomenological behavior, and we review all existing local experiments that could constrain the coupling constants in such a model.

Two different types of experiments are considered in this publication: (i) experiments based on the local comparison of clocks (and more generally on the local comparison of atomic sensors, sensitive to different combinations of the constants of Nature) and (ii) local measurement of the differential acceleration between two bodies of different compositions and located at the same position in space-time. 

Regarding the linear case, the scalar field is the sum of an oscillating contribution and a Yukawa contribution. The oscillating contribution can be identified as DM while the Yukawa interaction leads to a ``standard'' fifth interaction between bodies. This means that both types of signatures can be searched for using atomic sensors and UFF measurements. It turns out that atomic sensors are more sensitive to the oscillations while the UFF experiments are more sensitive to the Yukawa interaction. Existing results using local frequency comparisons between Cs and Rb hyperfine transitions using the dual atomic fountain from SYRTE~\cite{hees:2016uq} and using a local frequency comparison between two radio-frequency transition in two Dysprosium isotopes~\cite{van-tilburg:2015fj} and existing results on UFF measurements from the E\"ot-Wash group~\cite{smith:1999aa,schlamminger:2008zr,wagner:2012fk} and from the MICROSCOPE space mission~\cite{touboul:2017aa} are presented in Fig.~\ref{fig:lin} (see also~\cite{berge:2018aa}). For small masses of the scalar field, clock comparisons are the most constraining observations while for large masses, UFF measurements are more powerful. It is also interesting to mention that for intermediate masses, the result from the space experiment MICROSCOPE is more constraining than those from the E\"ot-Wash group (this is due to the sensitivity of the measurements), while for very large masses (typically for masses corresponding to a Yukawa interaction length $\lambda_\varphi < R_\oplus$), experiments located at the Earth's surface are more sensitive than experiments located in space. Finally, it is worth mentioning that so-called ``natural'' couplings (i.e. coupling coefficients $d_i^{(1)}$ of the order of unity) are excluded up to a scalar field mass of $\sim 10^{-5}$ eV (for $d_e^{(1)}$).

The case of the quadratic coupling is more complex and leads to a richer phenomenology. First of all, at the classical level, the solution for the scalar field shows that no Yukawa interaction is generated in this model\footnote{Quantum one-loop corrections are expected to produce an additional $1/r^3$ potential at distances $2m_\varphi r\ll 1$, see e.g. \cite{olive:2008fk}. This will not impact our results since these modifications are generally orders of magnitude smaller.}. Instead, the scalar field exhibits an oscillatory behavior that is perturbed or enhanced by the presence of a massive body. This important result is in agreement with \cite{de-pirey-saint-Alby:2017aa}. In addition, in Sec.~\ref{sec:sol_quad}, we show that the amplitude of the scalar field oscillations can be amplified (in the case of negative coupling coefficients $d_i^{(2)}$) or  screened (in the case of positive coupling coefficients $d_i^{(2)}$), a behavior similar to the scalarization mechanism~\cite{damour:1993vn,damour:1996uq}. In particular, it can be shown that for large positive coupling coefficients, the scalar field tends to vanish inside and at the surface of the central body. This has a direct consequence: experiments in space are more interesting to detect or constrain DM with a quadratic coupling to the standard model fields. The signature from such a scalar field on the comparison of two atomic sensors located at the same position takes the form of a constant that depends on the location, and of an oscillation whose amplitude depends on the location. Such a signature is completely new, has never been searched for in the past and favors experiments that would compare clocks located in eccentric orbits in space. Comparisons between frequencies on Earth provide interesting constraints on the coupling coefficients (see Fig.~\ref{fig:quad}), but they are somewhat limited because of the deamplification mechanism mentioned above. Regarding UFF measurements, two types of signatures can be searched for: a signature that corresponds exactly to the regular definition of the UFF $\eta$ parameter, and an oscillating signature. Except for very small scalar field masses, constraints from the UFF measurements are more powerful, and space experiments are more adapted to search for this type of DM candidate. The amplification and screening mechanisms lead to a modification of the constraints compared to what has been previously published in the literature.  Finally we point out that in the quadratic case so-called ``natural'' couplings are still allowed in most of the parameter space we explored.

This work is only a first step in the exploration of the signatures produced by scalar ultralight DM on some local experiments. In this paper, we focused on experiments that probe directly the Einstein Equivalence Principle. Further exploration is needed to consider experiments that are probing space-time curvature and not only the Einstein Equivalence Principle, like, e.g. the orbital dynamics using planetary ephemerides or Lunar Laser Ranging or the motion of S-stars around our Galactic Center~\cite{hees:2017aa,boskovic:2018aa}, light deflection, binary pulsars~\cite{khmelnitsky:2014aa,porayko:2014aa,blas:2017aa}, gravitational wave interferometers~\cite{stadnik:2015aa,stadnik:2016aa}, cosmological measurements like the Big Bang Nucleosynthesis~\cite{santiago:1997aa,damour:1999aa,copi:2004aa,cyburt:2005rz,coc:2006aa,larena:2007aa,iocco:2009la,stadnik:2015yu}, gravimetry \cite{geraci:2016aa}. For these experiments, one needs to solve the equations of the space-time metric in addition to the scalar field, whose solution is presented in Sec.~\ref{sec:phi}.

\begin{acknowledgments}
Y. V. S. was supported by the Humboldt Research Fellowship and gratefully acknowledges helpful discussions with Victor Flambaum.
\end{acknowledgments}

\bibliography{OM_add,../../../jpl/jpl_byme/biblio}

\appendix
\section{Other conventions used in the literature}\label{app:conventions}
Different conventions for the scalar field and for the coupling constants are used in the literature. In particular, while a dimensionless scalar field $\varphi$ is used in this paper as it is done in \cite{damour:2010ve,damour:2010zr}, a dimensionful scalar field $\phi$ is sometimes used \cite{arvanitaki:2015qy,arvanitaki:2016qv,stadnik:2015aa,stadnik:2015nr}. The action using this scalar field writes
\begin{equation}
	S=\frac{1}{c}\int d^4x \sqrt{-g}\left[\frac{R}{2\kappa} -\frac{1}{2}g^{\mu\nu}\partial_\mu\phi\partial_\nu\phi-\frac{m_\phi^2\phi^2}{2}\right] +S_\textrm{mat}\,. 
\end{equation}
Comparing this action with Eq.~(\ref{eq:action}) shows that
\begin{equation}
	\varphi=\left(4\pi G/c\hbar\right)^{1/2}\phi=\sqrt{4\pi}\phi/M_\textrm{Pl}\, ,
\end{equation}
with $M_\textrm{Pl}$ the Planck mass ($M_\textrm{Pl}=1.22\times 10^{19}$ GeV).

On the other hand, the coupling between the scalar field and matter is sometimes expressed using other conventions. This work uses dimensionless coefficients as used in \cite{damour:2010ve,damour:2010zr,arvanitaki:2015qy,arvanitaki:2016qv} while \cite{stadnik:2015aa,stadnik:2015nr,stadnik:2015yu,stadnik:2016aa,stadnik:2016kq,stadnik:2018aa,kalaydzhyan:2017aa} use coefficients that have the dimension of energy.

The  convention used by \cite{stadnik:2015aa,stadnik:2015nr,stadnik:2015yu,stadnik:2016aa,stadnik:2016kq,stadnik:2018aa,kalaydzhyan:2017aa} in the case of a linear coupling leads to 
\begin{subequations}
	\begin{align}
		\alpha_\textrm{EM}&=\alpha_\textrm{EM}\left(1+\frac{\phi}{\Lambda_\gamma}\right) \,  \\
		m_j&=m_j\left(1+\frac{\phi}{\Lambda_j}\right) \, , \quad \textrm{for }j=e,u,d \, ,
	\end{align}
\end{subequations}
for the linear case. A direct comparison with Eq.~(\ref{eq:constants}) leads to the following relation between the two sets of coefficients:
\begin{subequations}
	\begin{align}
		\Lambda_\gamma&=\frac{M_\textrm{Pl}}{\sqrt{4\pi}d_e^{(1)}} \, \\
		\Lambda_q &=\frac{M_\textrm{Pl}}{\sqrt{4\pi}d_{\hat m}^{(1)}}\, \\
		\Lambda_e &=\frac{M_\textrm{Pl}}{\sqrt{4\pi}d_{m_e}^{(1)}}\, .
	\end{align}
\end{subequations}

On the other hand, the convention used by \cite{stadnik:2015aa,stadnik:2015nr,stadnik:2015yu,stadnik:2016aa,stadnik:2016kq,stadnik:2018aa,kalaydzhyan:2017aa} in the case of a quadratic coupling leads to
\begin{subequations}
	\begin{align}
		\alpha_\textrm{EM}&=\alpha_\textrm{EM}\left(1+\frac{\phi^2}{\left(\Lambda'_\gamma\right)^2}\right) \,  \\
		m_j&=m_j\left(1+\frac{\phi}{\left(\Lambda'_j\right)^2}\right) \, , \quad \textrm{for }j=e,u,d \, ,
	\end{align}
\end{subequations}

\begin{subequations}
	\begin{align}
		\Lambda'_\gamma&=\frac{M_\textrm{Pl}}{\sqrt{2\pi d_e^{(2)}}} \, \\
		\Lambda'_q &=\frac{M_\textrm{Pl}}{\sqrt{2\pi d_{\hat m}^{(2)}}}\, \\
		\Lambda'_e &=\frac{M_\textrm{Pl}}{\sqrt{2\pi d_{m_e}^{(2)}}}\, .
	\end{align}
\end{subequations}

\section{Dilaton charges}\label{app:dilatonic_charges}
In this Appendix, we briefly remind the formulas that were used to get the values in Tab.~\ref{tab:dilaton_coef}. These formulas are derived in \cite{damour:2010zr}. From \cite{damour:2010zr,damour:2010zr}, one can write the coupling 
\begin{align}
	\tilde\alpha^{(i)}_A =&d_g^{(i)}+\left[Q_{\hat m}\right]_A\Big(d_{\hat m}^{(i)}-d_g^{(i)}\Big) +\left[Q_{\delta m}\right]_A\Big(d_{\delta m}^{(i)}-d_g^{(i)}\Big) \nonumber\\
	&+\left[Q_{ m_e}\right]_A\Big(d_{ m_e}^{(i)}-d_g^{(i)}\Big)+ \left[Q_{e}\right]_Ad_e^{(i)}\, ,
\end{align}
where the dilaton charges write
\begin{subequations}
	\begin{align}
		Q_{\hat m}=&F_A\Bigg[0.093-\frac{0.036}{A^{1/3}}-0.02\frac{(A-2Z)^2}{A^2}\nonumber\\
		&\qquad -1.4\times 10^{-4}\frac{Z(Z-1)}{A^{4/3}}\Bigg]\\
		Q_{\delta m}=&F_A\Bigg[0.0017\frac{A-2Z}{A}\Bigg] \\
		Q_{m_e}=&F_A\Bigg[5.5\times 10^{-4}\frac{Z}{A}\Bigg]\\
		Q_e=&F_A\Bigg[-1.4+8.2 \frac{Z}{A}+7.7 \frac{Z(Z-1)}{A^{4/3}}\Bigg]\times 10^{-4}\, ,
	\end{align}
	with
	\begin{equation}
		F_A=\frac{Am_\textrm{amu}}{m_A}= 1+\mathcal O\left(10^{-4}\right)\, ,
	\end{equation}
	where $Z$ is the atomic number, $A$ is the mass number, $m_\textrm{amu}=931$ MeV and $m_A$ is the mass of the atom.
\end{subequations}
It is convenient to decompose $\tilde\alpha^{(i)}_A$ into a composition independent part and a part that is composition dependent and that will play an important role in EEP tests. In order to provide such a decomposition, we will use the fact that $Z/A\sim 1/2$ for most (heavy) elements to get
\begin{equation}
	\tilde \alpha^{(i)}_A=d_g^{*(i)} + \bar\alpha^{(i)}_A \, ,
\end{equation}
where $d_g^{*(i)}$ contains the composition independent parts of the dilatonic charges
\begin{align}
	d_g^{*(i)}=&d_g^{(i)}+0.093\Big(d_{\hat m}^{(i)}-d_g^{(i)}\Big)+2.75\times 10^{-4}\Big(d_{ m_e}^{(i)}-d_g^{(i)}\Big)\nonumber\\
	&\qquad\quad+2.7\times 10^{-4}d_e^{(i)}\, ,
\end{align}
and where the composition dependent part of the coupling writes
\begin{align}
	\bar\alpha^{(i)}=&\left[Q'_{\hat m}\right]_A\Big(d_{\hat m}^{(i)}-d_g^{(i)}\Big) +\left[Q'_{\delta m}\right]_A\Big(d_{\delta m}^{(i)}-d_g^{(i)}\Big) \nonumber\\
	&+\left[Q'_{ m_e}\right]_A\Big(d_{ m_e}^{(i)}-d_g^{(i)}\Big)+ \left[Q'_{e}\right]_Ad_e^{(i)}\, .
\end{align}
These new dilatonic charges are now given by
\begin{subequations}\label{eq:Q'}
	\begin{align}
		Q'_{\hat m}&=-\frac{0.036}{A^{1/3}}-0.02\frac{(A-2Z)^2}{A^2} -1.4\times 10^{-4}\frac{Z(Z-1)}{A^{4/3}}\\
		Q'_{\delta m}&= 0.0017\frac{A-2Z}{A} \\
		Q'_{m_e}&=-2.75\times 10^{-4}\frac{A-2Z}{A}\\
		Q'_e&=-4.1\times 10^{-4} \frac{A-2Z}{A}+7.7\times 10^{-4} \frac{Z(Z-1)}{A^{4/3}}\, .	
	\end{align}
\end{subequations}
In the previous equations, the terms that are proportional to $(A-2Z)^n$ are usually negligible for heavy elements.

\section{Solutions for the scalar field}
\label{app:sol}

\subsection{Linear coupling}
The equation for the scalar field is given by
\begin{equation}
	\frac{1}{c^2}\ddot \varphi(t,\bm x) - \Delta \varphi(t,\bm x) =  -\frac{4\pi G }{c^2}\tilde\alpha^{(1)}_A \rho_A(\bm x)-\frac{c^2m_\varphi^2}{\hbar^2}\varphi(t,\bm x) \label{eq:KG_lin} \, .
\end{equation}
The general solution of this equation is the sum of the general solution of the homogeneous equation and of a particular solution. The homogeneous equation is a regular wave equation whose solutions are plane-waves $\varphi_0 \cos \left(\bm k .\bm x -\omega t+\delta\right)$ where  $\left|\bm k\right|^2 + c^2 m_\varphi^2/\hbar^2=\omega^2/c^2$.

A particular solution of the non-homogeneous equation can be obtained by considering that the source term depends only on the spatial coordinates ($\rho$ is time independent). The particular solution will therefore be time independent and can be determined by computing the Green's function $G(\bm x)$, solution of the equation
\begin{equation}
	 \Delta G(\bm x) -\frac{c^2m_\varphi^2}{\hbar^2}G(\bm x)=\delta^{(3)}(\bm x)\, .
\end{equation}
The isotropic solution is given by
\begin{equation}
	G(\bm x)=-\frac{1}{4\pi}\frac{e^{-r/\lambda_\varphi}}{r}\, ,
\end{equation}
with $\lambda_\varphi=\hbar/cm_\varphi$. The general solution for the scalar field is therefore given by 
\begin{align}
	\varphi^{(1)}(t,\bm x) &=\varphi_0 \cos \left(\bm k .\bm x -\omega t+\delta\right)\label{eq:varphi1_ap} \\
	&- \frac{G}{c^2} \int d^3\bm x' \frac{e^{-\left|\bm x-\bm x'\right|/\lambda_\varphi}}{\left|\bm x-\bm x'\right|}\tilde\alpha^{(1)}(\bm x')\rho(\bm x') \, .\nonumber
\end{align}

\subsubsection{Test mass}
The academic case of a test mass whose density is given by $\rho(\bm x)=M_A \delta(\bm x)$ leads to the standard Yukawa form of the part of the scalar field generated by the body $A$
\begin{equation}
	\varphi^{(1)}_A(\bm x) = -\tilde \alpha_A^{(1)}\frac{GM_A}{rc^2}e^{-r /\lambda_\varphi}\, .
\end{equation}

\subsubsection{Homogeneous spherically symmetric body}
If we consider a uniform extended spherically symmetric body, characterized by $\rho(\bm x)=\rho_A$ if $r < R_A$ and 0 otherwise with $\rho_A=3M_A/4\pi R_A^3$, the integration from Eq.~(\ref{eq:varphi1_ap}) leads to~\cite{adelberger:2003uq}
\begin{equation}
	\varphi^{(1)}_A(\bm x) = -\tilde \alpha_A^{(1)}I\left(\frac{R_A}{\lambda_\varphi}\right)\frac{GM_A}{rc^2}e^{-r /\lambda_\varphi}\, ,
\end{equation}
with
\begin{equation}
	I(x) =3 \frac{x\cosh x-\sinh x}{x^3}\, .
\end{equation}

\subsubsection{Two-layer spherically symmetric body}
Let's now consider a spherically symmetric body composed of two layers (like e.g. the Earth with a core and a mantle). The matter density is given by
\begin{align}
	\rho(\bm x) &= \rho_1 \phantom{0}\quad \textrm{if} \, r \leq R_1 \, ,\\
	& =  \rho_2 \phantom{0}\quad \textrm{if} \, R_1 < r \leq R_2 \, ,\\
	&= 0 \phantom{\rho_1}\quad \textrm{if} \, R_2 < r \, .
\end{align}
The coupling constant $\tilde \alpha^{(1)}$ is also dependent on the position 
\begin{align}
	\tilde \alpha^{(1)}(\bm x) &= \tilde \alpha^{(1)}_1 \phantom{0}\quad \textrm{if} \, r \leq R_1 \, ,\\
	& =  \tilde \alpha^{(1)}_2 \phantom{0}\quad \textrm{if} \, R_1 < r \leq R_2 \, ,\\
	&= 0 \phantom{\tilde \alpha^{(1)}_2}\quad \textrm{if} \, R_2 < r \, .
\end{align}
The integration of Eq.~(\ref{eq:varphi1_ap}) gives 
\begin{align}
	\varphi^{(1)}_A(\bm x) &= - \frac{GM}{c^2r}e^{-r/\lambda_\varphi}\Bigg[\tilde\alpha^{(1)}_1 \frac{M_1}{M}I\left(\frac{R_1}{\lambda_\varphi}\right)\\
	     &\qquad + \tilde\alpha^{(1)}_2 \frac{M_2}{M} \, \frac{R_2^3 I\left(\frac{R_2}{\lambda_\varphi}\right)-R_1^3I\left(\frac{R_1}{\lambda_\varphi}\right)}{R_2^3-R_1^3}\Bigg] \, ,\nonumber 
\end{align}
where $M_1=4\pi R_1^3\rho_1/3$ is the mass of the internal core, $M_2=4 \pi \rho_2 (R_2^3-R_1^3)/3$ is the mass of the external shell and $M=M_1+M_2$.

\subsubsection{Summary}
To summarize, the general solution for the scalar field around a spherically symmetric body is given by
\begin{align}\label{eq:phi_1_ap}
	\varphi^{(1)}(t,\bm x) &=  \varphi_0 \cos \left(\bm k .\bm x -\omega t+\delta\right) - s^{(1)}_A \frac{GM_A}{c^2r}e^{-r/\lambda_\varphi}\, , 
\end{align}
where
\begin{subequations}\label{eq:s1_app}
\begin{align}
	s^{(1)}_A =& \tilde\alpha_A^{(1)} \quad \textrm{for a point particle}\, , \\
	 =&\tilde \alpha_A^{(1)}I\left(\frac{R_A}{\lambda_\varphi}\right) \quad \textrm{for a sphere} \, ,\\
	=&\tilde\alpha^{(1)}_2 \frac{M_2}{M} \, \frac{R_2^3 I\left(\frac{R_2}{\lambda_\varphi}\right)-R_1^3I\left(\frac{R_1}{\lambda_\varphi}\right)}{R_2^3-R_1^3} \\
	&+\tilde\alpha^{(1)}_1 \frac{M_1}{M}I\left(\frac{R_1}{\lambda_\varphi}\right) \quad \textrm{for a two-layer sphere} \, .\nonumber
\end{align}
\end{subequations}

\subsection{Quadratic coupling}
The equation for the scalar field is given by
\begin{equation}
	\frac{1}{c^2}\ddot \varphi(t,\bm x) - \Delta \varphi(t,\bm x) =  -\frac{4\pi G }{c^2}\tilde\alpha^{(2)}_A \varphi(t,\bm x) \rho_A(\bm x)-\frac{c^2m_\varphi^2}{\hbar^2}\varphi(t,\bm x) \label{eq:KG_quad} \, .
\end{equation}
This is a fully linear equation with no source term. The trivial solution $\varphi=0$ is always a solution to this equation.
In order to find a non-trivial solution, let us use a separation of variables and write the scalar field as a product of two functions (one time dependent and one space dependent)
\begin{equation}
	\varphi(t,\bm x)=T(t)\, X(\bm x) \,.
\end{equation}
Using this ansatz, Eq.~(\ref{eq:KG_quad}) is therefore equivalent to 
\begin{subequations}
\begin{align}
	\ddot T-\alpha T &= 0\, \\
	\Delta X  + \beta X  - \frac{4\pi G}{c^2} \tilde\alpha^{(2)}(\bm x)\rho(\bm x)X &=0\label{eq:X}\\
	\alpha+\beta &= m_\varphi^2 \, .
\end{align}
\end{subequations}

We are interested in finding the solution in the case around a spherically symmetric body. The outside solution for the function $X$ is a solution of 
\begin{equation}
	\Delta X  + \beta X   =0\, .
\end{equation}
This equation presents different behavior depending on the value of $\beta$. Since we want to identify the scalar field as DM, we are interested in the solutions that remain non-vanishing at infinity (and that remain finite). This behavior only shows up for $\beta=0$, which will be considered hereafter. In that case, the temporal part of the scalar field can be solved easily and 
\begin{equation}
\label{eqapp:varphiansatz}
	\varphi(t,\bm x)=\varphi_0 \cos(\omega t +\delta ) X(\bm x) \,,
\end{equation}
with $\omega=m_\varphi c^2/\hbar$.

The function $X(\bm x)$ depends on the specific modeling of the body although, the form of the outside solution will be 
\begin{equation}\label{eq:extern}
	X(\bm x)=1 + \frac{A}{r} \, ,
\end{equation}
where the constant $A$ will depend on the internal structure of the central body. We will model the central body as an extended spherical mass or as a two-layer sphere. In both cases, the strategy to solve for the function $X(\bm x)$ is to solve Eq.~(\ref{eq:X}) in each layer, to keep solutions that remains finite for $r=0$ and whose radial derivative at $r=0$ vanishes and to apply continuity conditions (continuity of the scalar field and of its derivative) at the interfaces.

\subsubsection{Homogeneous spherically symmetric body}\label{ap:scal_quad}
If we consider a uniform extended spherically symmetric body, characterized by $\rho(\bm x)=\rho_A$ if $r < R_A$ and 0 otherwise with $\rho_A=3M_A/4\pi R_A^3$, inside the body $X(\bm x)=X(r)$ is a solution of
\begin{equation}
	\Delta X = \frac{4\pi G}{c^2} \tilde\alpha^{(2)}_A \rho_A X \, .
\end{equation}
The solution that remains finite and whose radial derivative vanishes at $r=0$ is given by
\begin{subequations}\label{eq:intern}
\begin{align}
	X(r) & = B \frac{\sinh \gamma_A r}{r} \quad \textrm{if} \quad \tilde\alpha^{(2)}_A>0 \, \\
	     & = B \frac{\sin \gamma_A r}{r} \quad \textrm{if} \quad \tilde\alpha^{(2)}_A<0 \,  ,
\end{align}
\end{subequations}
where
\begin{equation}
	\gamma_A^2 = \frac{4\pi G}{c^2} \left|\tilde\alpha^{(2)}_A \right| \rho_A =3 \left|\tilde\alpha^{(2)}_A \right| \frac{GM_A}{c^2R_A^3}\, .
\end{equation}
The continuity conditions (continuity of $X$ and of its radial derivative) at the interface $r=R_A$ between the interior solution from Eq.~(\ref{eq:intern}) and the exterior solution from Eq.~(\ref{eq:extern}) allows one to determine the constants $A$ and $B$.

The final solution depends on the sign of $\tilde\alpha_A^{(2)}$ and is given by
\begin{align}
	X(r)&= K_{\textrm{sign}[\tilde \alpha_A^{(2)}]} \left(\frac{r}{R_A},\sqrt{3\left|\tilde \alpha_A^{(2)}\right| \frac{GM_A}{c^2R_A}}\right)\quad \textrm{for} \, r\leq R_A \label{eqapp:solX} \\
	 & =1 -\tilde\alpha_A^{(2)}\frac{GM_A}{c^2r} J_{\textrm{sign}[\tilde \alpha_A^{(2)}]} \left(\sqrt{3\left|\tilde \alpha_A^{(2)}\right| \frac{GM_A}{c^2R_A}}\right) \quad \textrm{for} \, r> R_A \, ,\nonumber
\end{align}
with
\begin{subequations}\label{eq:J}
	\begin{align}
		J_+(x)&=3\frac{x-\tanh x}{x^3} \\
		J_-(x)&=3\frac{\tan x-x}{x^3} \\
		K_+(x,y)&= \frac{\textrm{sinhc}(xy)}{\cosh(y)}\label{eqapp:Kp} \\
		K_-(x,y)&= \frac{\textrm{sinc}(x y)}{\cos (y)}\, .
	\end{align}
\end{subequations}

\subsubsection{Two-layer spherically symmetric body}
Let's now consider a spherically symmetric body composed of two layers (like e.g. the Earth with a core and a mantle). The matter density is given by
\begin{align}
	\rho(\bm x) &= \rho_1 \phantom{0}\quad \textrm{if} \, r \leq R_1 \, ,\\
	& =  \rho_2 \phantom{0}\quad \textrm{if} \, R_1 < r \leq R_2 \, ,\\
	&= 0 \phantom{\rho_1}\quad \textrm{if} \, R_2 < r \, .
\end{align}
The coupling constant $\tilde \alpha^{(2)}$ is also dependent on the position 
\begin{align}
	\tilde \alpha^{(2)}(\bm x) &= \tilde \alpha^{(2)}_1 \phantom{0}\quad \textrm{if} \, r \leq R_1 \, ,\\
	& =  \tilde \alpha^{(2)}_2 \phantom{0}\quad \textrm{if} \, R_1 < r \leq R_2 \, ,\\
	&= 0 \phantom{\tilde \alpha^{(2)}_2}\quad \textrm{if} \, R_2 < r \, .
\end{align}

The solution outside the body is given by Eq.~(\ref{eq:extern}), the solution within the first layer depends on the sign of $\tilde \alpha^{(2)}_1$ and is given by Eq.~(\ref{eq:intern}) while the solution within the external layer is given by
\begin{align}
	X(r) &= C\frac{e^{\gamma_2 r}}{r} + D\frac{e^{-\gamma_2 r}}{r} \quad \textrm{if} \quad \tilde\alpha^{(2)}_2>0 \, \\
	&= C\frac{\sin \gamma_2 r}{r} + D\frac{\cos \gamma_2 r}{r} \quad \textrm{if} \quad \tilde\alpha^{(2)}_2<0 \, , \nonumber
\end{align}
with
\begin{equation}\label{eq:gammai}
		\gamma_i^2 = \frac{4\pi G}{c^2} \left|\tilde\alpha^{(2)}_i \right| \rho_i \, .
\end{equation}
The continuity conditions at the two interfaces $r=R_1$ and $r=R_2$ allows one to determine the 4 constants $A$, $B$, $C$ and $D$. 
After solving this system of equations, the external solution is given by
\begin{align}
	X(r)= 1 - \frac{GM}{c^2r}L\left(\tilde\alpha^{(2)}_1,\tilde\alpha^{(2)}_2,R_1,R_2,\rho_1,\rho_2\right)\, ,
\end{align}
where the function $L$ is given by
\begin{widetext}
\begin{subequations}\label{eq:L}
	\begin{align}	L&=\frac{\cosh\left[(R_1-R_2)\gamma_2\right]\Bigg(R_2\gamma_1\cosh\left[R_1\gamma_1\right]-\sinh\left[R_1\gamma_1\right]\Bigg)+\sinh\left[(R_1-R_2)\gamma_2\right]\Bigg(\frac{\gamma_1}{\gamma_2}\cosh\left[R_1\gamma_1\right]-R_2\gamma_2\sinh\left[R_1\gamma_1\right]\Bigg)}{\frac{GM}{c^2}\Bigg(\gamma_2\sinh\left[R_1\gamma_1\right]\sinh\left[(R_1-R_2)\gamma_2\right]-\gamma_1\cosh\left[R_1\gamma_1\right]\cosh\left[(R_1-R_2)\gamma_2\right]\Bigg)} \\
		& \hspace{13cm}\textrm{if}\quad \alpha^{(2)}_i >0 \nonumber\\
	&=\frac{\cos\left[(R_1-R_2)\gamma_2\right]\Bigg(\sin\left[R_1\gamma_1\right]-R_2\gamma_1\cos\left[R_1\gamma_1\right]\Bigg)-\sin\left[(R_1-R_2)\gamma_2\right]\Bigg(\frac{\gamma_1}{\gamma_2}\cos\left[R_1\gamma_1\right]+R_2\gamma_2\sin\left[R_1\gamma_1\right]\Bigg)}{\frac{GM}{c^2}\Bigg(\gamma_2\sin\left[R_1\gamma_1\right]\sin\left[(R_1-R_2)\gamma_2\right]+\gamma_1\cos\left[R_1\gamma_1\right]\cos\left[(R_1-R_2)\gamma_2\right]\Bigg)} \\
		& \hspace{13cm}\textrm{if} \quad \alpha^{(2)}_i <0\, , \nonumber
	\end{align}
\end{subequations}
\end{widetext}
where the $\gamma_i$ are given by Eq.~(\ref{eq:gammai}) and $M=\frac{4}{3}\pi R_1^3\rho_1 + \frac{4}{3}\pi \rho_2 \left(R_2^3-R_1^3\right)$.

\subsubsection{Summary}
To summarize, the general solution for the scalar field around a spherically symmetric body that is not vanishing at infinity is given by
\begin{equation}
	\varphi^{(2)}=\varphi_0\cos\left(\frac{m_\varphi c^2}{\hbar}t+\delta\right)\left[1-s_A^{(2)}\frac{GM_A}{c^2r}\right]\, ,
\end{equation}
where the scalar charge $s_A^{(2)}$ is given by
\begin{subequations}
\begin{align}
	s_A^{(2)}&=\tilde\alpha_A^{(2)} J_{\textrm{sign}[\tilde \alpha_A^{(2)}]} \left(\sqrt{3\left|\tilde \alpha_A^{(2)}\right| \frac{GM_A}{c^2R_A}}\right)
\end{align}
for an extended homogeneous spherically symmetric body and
\begin{align}
	s_A^{(2)}&=L\left(\tilde\alpha^{(2)}_1,\tilde\alpha^{(2)}_2,R_1,R_2,\rho_1,\rho_2\right)
\end{align}
\end{subequations}
for a two-layer body where the function $J$ is defined by Eq.~(\ref{eq:J}) and the function $L$ is defined by Eq.~(\ref{eq:L})
\end{document}